\documentclass[
reprint,
 amsmath,amssymb,bm,
pre
floatfix,
]{revtex4-2}

\usepackage{lipsum}
\usepackage{graphicx}
\usepackage{dcolumn}
\usepackage{bm}
\usepackage{appendix} 
\usepackage{physics}
\usepackage{algorithm}
\usepackage{algorithmic}
\usepackage{subfigure}
\usepackage{graphicx}
\usepackage{color}

\usepackage{hyperref}

\begin{document}

\title{From a feedback-controlled demon to an information ratchet in a double quantum dot}


\author{Debankur Bhattacharyya}
\email{dbhattac@umd.edu}
\affiliation{Institute for Physical Science and Technology, University of Maryland, College Park, Maryland 20742, USA}
\author{Christopher Jarzynski}%
\email{cjarzyns@umd.edu}
\affiliation{Institute for Physical Science and Technology, Department of Chemistry and Biochemistry,\\ and Department of Physics, University of Maryland, College Park, Maryland 20742, USA}





\date{\today}

\begin{abstract}
We present a simple strategy for constructing an information ratchet or memory-tape model of Maxwell's demon, from a feedback-controlled model. We illustrate our approach by converting the Annby-Andersson feedback-controlled double quantum dot model [\textit{Phys.  Rev.  B} \textbf{101},165404 (2020)] to a memory-tape model. We use the underlying network structure of the original model to design a set of bit interaction rules for the information ratchet. The new model is solved analytically in the limit of long interaction times. For finite-time interactions, semi-analytical phase diagrams of operational modes are obtained. Stochastic simulations are used to verify theoretical results.
\end{abstract}

\maketitle


\section{Introduction}

 \begin{figure*}
    \centering
    \includegraphics[scale=0.35]{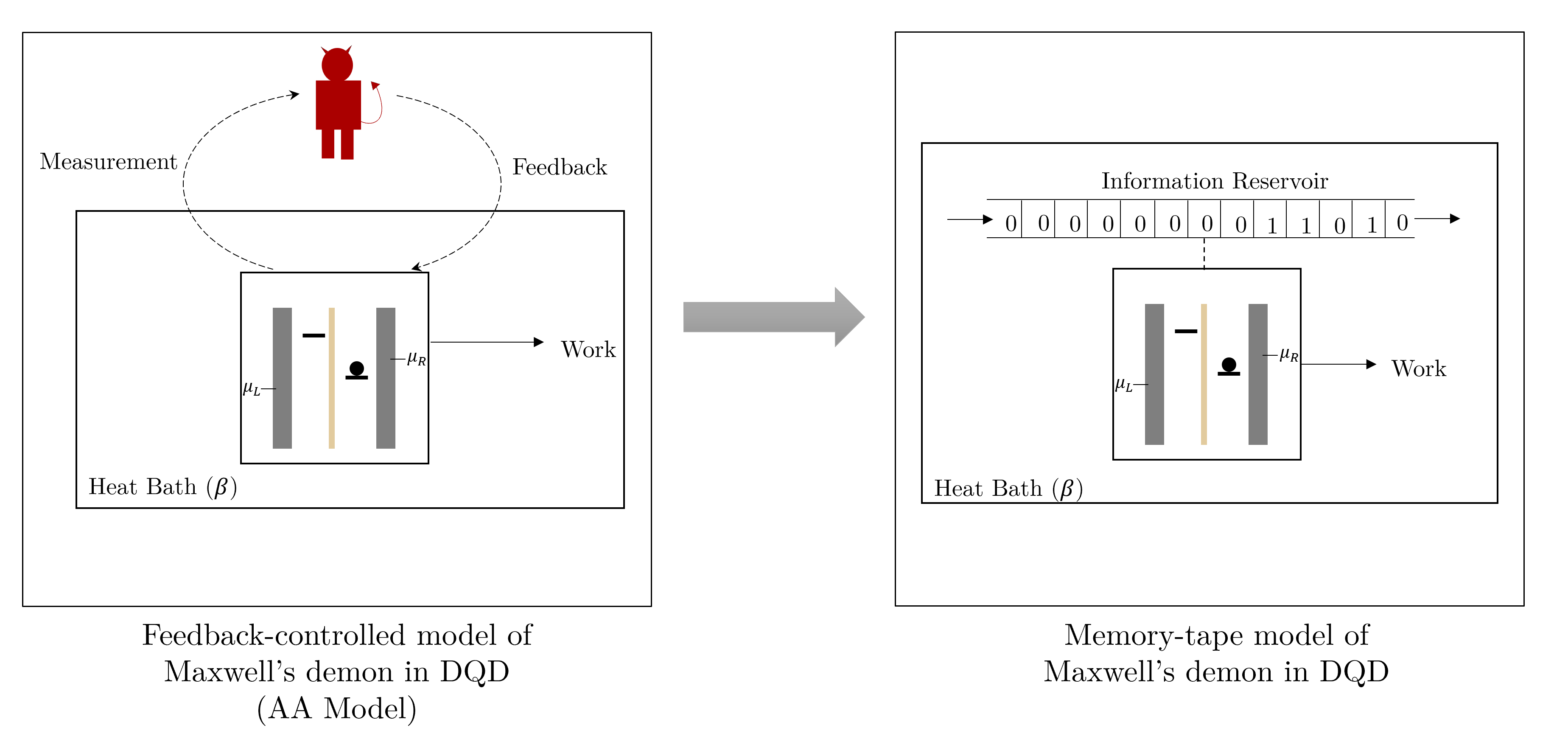}
    \caption{Two paradigms of Maxwell's demon. The left figure depicts the AA model \cite{AA_model}, a feedback-controlled model. On the right we show the corresponding memory-tape model or information ratchet. In both cases, heat from a thermal reservoir is converted to work, either through measurement and feedback, or through interaction with an information reservoir. We explore a strategy to convert a feedback-controlled model to a memory-tape model.}
    \label{fig:graphical_abstract}
\end{figure*}
In a letter to Peter Tait in 1867, James Maxwell described a thought experiment in which a hypothetical creature – now known as Maxwell’s demon – seemingly beats the second law of thermodynamics by performing measurements and feedback on the motions of individual gas molecules. 
This thought experiment initiated a line of research that exposed deep connections between thermodynamics and information processing.
Recent years have seen renewed activity in this field, with theoretical progress as well as experimental studies adding to our understanding of these connections.
See for example Ref \cite{Rex_review} for a review of the history of Maxwell’s demon and the current experimental state of the art, and Ref \cite{Parrando_review} for an introduction to the theoretical formalism of information thermodynamics.\par 

Over the past century and a half, numerous variations of Maxwell’s demon have been proposed, notably including Smoluchowski’s trapdoor \cite{Smoluchowski1927,planck1914vortrage,Skordos_comp_smoluchowski_trapdoor}, Szilard’s engine \cite{SzilardEngine}, and Feynman’s ratchet-and-pawl \cite{feynman2011feynman,REIMANN200257} as early examples.
These and other demons can broadly be classified according to two paradigms: feedback-controlled demons and autonomous demons. In the feedback-controlled paradigm, exemplified by both Maxwell’s thought experiment and Szilard’s engine, the demon is an external agent who performs measurements and provides feedback based on the outcomes of those measurements.
The physical nature of the demon is unspecified and irrelevant. Recent stochastic-thermodynamic analyses of feedback-controlled demons have led to the discovery of inequalities and fluctuation relations that sharpen our understanding of information thermodynamics. \cite{esposito2012stochastic,Sagawa_Ueda_FT_PRL_2012,Sagawa_Ueda_PRE_2012,lahiri2012fluctuation,Suri_Horowitz_FT_PRE_2010,Potts_Samuelsson_FT_PRL_2018,Zeng_Wang_FT_2021,Seifert_FT_review_2012,Sagawa_PRL_causal_networks,Shiraishi_2015_New_J_Phys} \par

In the autonomous paradigm, by contrast, the demon is a physical device that is cleverly designed to achieve, through the rectification of thermal noise, the same outcome that a feedback-controlled demon achieves through measurements and feedback.  Smoluchowski’s trapdoor and Feynman’s ratchet-and-pawl are examples of autonomous demons that at first glance appear to bring about their intended outcomes, but upon closer inspection are seen to fail.
These examples are often held up as cautionary tales highlighting the futility of trying to invent a gadget that defies the second law of thermodynamics.
Bennett, however, argued that an autonomous demon could succeed if it were coupled to a memory storage device \cite{Bennett1973,Bennet1982,BENNETT2003}. 
Due to the physical nature of information, as proposed by Landauer \cite{LandauerLimit,Landauer1991}, these models of autonomous demons can utilize the memory storage device as an \textit{information reservoir} \cite{Deffner_CJ_PRX_2013} to mimic the same functionality as feedback-controlled models of Maxwell's demon.
In these models, the decrease of environmental entropy is compensated by an increase in the randomness of the informational state of the memory storage device. Often these models of autonomous Maxwell's demons or \textit{information ratchets} are visualized by imagining a system that interacts with a sliding memory-tape containing a sequence of bits (the information reservoir) and are called \textit{memory-tape models}. \par
 Mandal and Jarzynski (MJ) introduced an explicit stochastic model of a memory-tape autonomous demon \cite{MJ_model}, and similar models have been developed for both classical and quantum systems \cite{MJ_refrigarator,Lu_mechanical_demon,Lu_programmable_demon,Barato_Seifert_EPL_2013,Barato_Seifert_PRL_2014,Barato_Seifert_PRE_2014,Engel_2014,InfoEngine_spatially_structured,Peng_Duan_2016,rana2016multipurpose,Strasberg_PRE_2014_spin_valve,HTQuan_PRE2015_enzyme,Deffner_quantum,Miyake_PRE_2015_quantum_demon,Deffner_PRE_2019_quantum}.
Memory-tape models have recently been studied as \textit{transducers} from the computer science and information theory perspective, and the thermodynamic implications of correlation among the bits of the memory-tape have been discussed \cite{Boyd2016,boyd2017leveraging,Boyd_IPSL_PRL_2017,Boyd_PRE_2017_correlation_powered_demon,Boyd_PRX_2018,Jurgens_PRR_2020,NeriMerhav2015,NeriMarhav_2017_IPSL,wolpert2019stochastic}. \textit{Repeated interaction models}, commonly used in quantum thermodynamics, have also been used to discuss the thermodynamics of the memory-tape models \cite{Strasberg_Repeated_interaction_PRX_2017}. Autonomous demons can be also be driven by temperature differences (see Refs.~\cite{Potts_PhysRevResearch_2019,Esposito_PhysRevE_2021} and references therein), but in the present work we focus on autonomous models driven by information reservoirs. \par
 
A number of authors have explored connections between feedback-controlled and memory-tape models \cite{AMF_Horowitz_2013,Barato_Seifert_PRL_2014,bauer2014optimized,shiraishi2016measurement,Strasberg_PRE_2014_spin_valve}. 
Horowitz \textit{et al} \cite{AMF_Horowitz_2013} designed a feedback-controlled information motor based on the system-bit interactions of Ref \cite{MJ_model}.
Barato and Seifert \cite{Barato_Seifert_PRE_2014,Barato_Seifert_PRL_2014} discussed a stochastic thermodynamics \cite{Seifert_FT_review_2012} framework that encompasses both feedback-controlled and memory-tape models. 
Shiraishi \textit{et al} \cite{shiraishi2016measurement} showed that the measurement-feedback model introduced in Ref. \cite{AMF_Horowitz_2013} can be reduced to the simplified MJ model of Ref. \cite{Barato_Seifert_EPL_2013,Barato_Seifert_PRE_2014,Barato_Seifert_PRL_2014}.
Strasberg \textit{et al} \cite{Strasberg_PRE_2014_spin_valve} described a system with a spin-valve and a quantum dot that can mimic the thermodynamic behaviour of the MJ model and can be mapped to a Brownian ratchet. They also presented a feedback-controlled model that captures the effective dynamics of the corresponding memory-tape model, and they compared how the second law of thermodynamics applies to these two paradigms. \par 

In this article we use a recently introduced model \cite{AA_model} to develop and illustrate a general strategy for converting a non-autonomous, feedback-controlled model of Maxwell's demon into an autonomous, memory-tape model, or information ratchet.
Our approach uses network based modelling \cite{Seifert_FT_review_2012,Bipartite_2014} of a system of master equations, originally introduced by Schnakenberg in Ref. \cite{Schnakenberg_RevModPhys_1976}, to show how a non-autonomous demon with a seemingly complicated feedback protocol can systematically be modified to construct a memory-tape model that mimics its behavior. We illustrate this strategy by applying it to the recently proposed \textit{Annby-Andersson} (AA) model \cite{AA_model} of a double quantum dot (DQD) \cite{DQD_RevModPhys.75.1,DQD_RevModPhys.79.1217}.
We then present a theoretical analysis of the resulting memory-tape model. Our model has distinct regions in parameter space where it operates either as an \textit{information engine} or as a \textit{Landauer's eraser}. We solve the model exactly in the limit when each bit interacts with the DQD for an infinite amount of time, obtaining analytic expressions for thermodynamic quantities and critical parameter values. We also semi-analytically explore the finite time bit-interaction situation and show the corresponding phase diagrams. Lastly, we discuss a scheme for the stochastic simulation for memory-tape models and use it to simulate our model to verify the semi-analytical results. We limit our discussion to a completely classical stochastic model and leave quantum models as a future avenue for research. \par 

This article is organized as follows. In Sec.~\ref{Sec: background_setup}, we briefly review the Annby-Andersson model \cite{AA_model}, which is the starting point for designing our memory-tape model. Details of network based stochastic modeling \cite{Schnakenberg_RevModPhys_1976,Bipartite_2014,Seifert_FT_review_2012} are presented in Sec.~\ref{Sec:Stochastic_Modelling}. In Sec~\ref{Sec: 9_states} we map the AA model to a nine-state network by converting its control parameter to a stochastic variable. In Sec.~\ref{Sec:bit_coupling} and Sec.~\ref{Sec: 18states}, we discuss how to couple the DQD with incoming bits to mimic the behavior of the feedback-controlled demon. A summary of the general modelling scheme is presented in Sec.~\ref{Sec:Summary_modelling_strategy}. The analysis of memory-tape models is discussed in Sec.~\ref{Sec:Analyses_methods} following Ref.~\cite{MJ_model}. In Sec.~\ref{Sec:Thermo} we discuss the thermodynamics of our model and solve for analytical expressions of thermodynamics quantities in Sec.~\ref{Sec:Analytical_results}. Phase diagrams of operational modes are discussed in Sec.~\ref{Sec:phase_diagram} and  the stochastic simulation scheme for the model is presented in Sec.~\ref{Sec:Stochastic_Sim}.
\begin{figure*}
    \centering
    \includegraphics[scale=0.45]{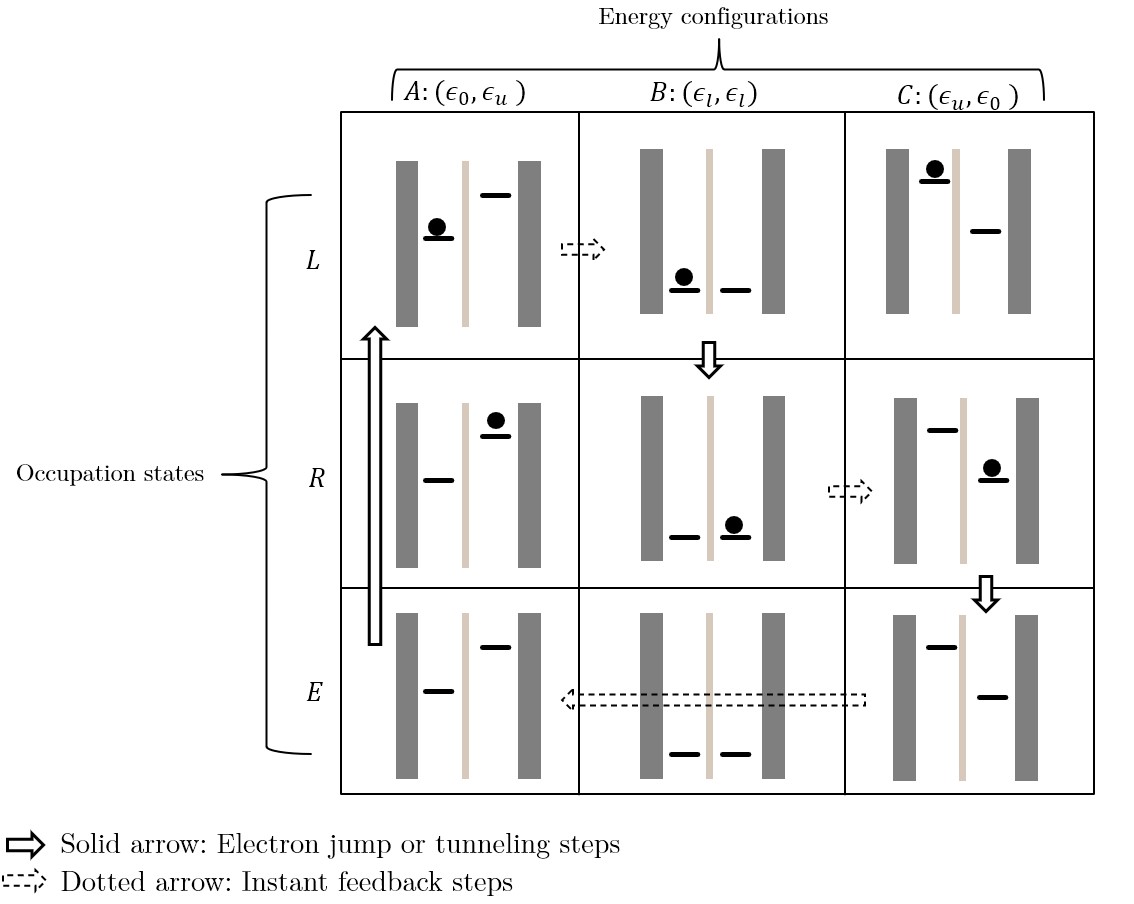}
    \caption{States of the double quantum dot system $x\equiv (\lambda, \sigma)$ and the protocol ($\mathcal{C}_{AA}$) for the AA model. Feedback steps (changes in $\lambda$) are shown using dotted arrows, and electron jumps (changes in $\sigma$) are shown using solid arrow.}
    \label{fig:DQD_states}
\end{figure*}

\section{Background and Setup: Annby-Andersson model}
\label{Sec: background_setup}
A quantum dot (QD) is an artificial nano-scale structure for confining electrons. Since it is possible to tune the energetic cost of adding an excess electron,  QDs act like `artificial atoms' with tunable energy levels. \cite{QD_ref,DQD_RevModPhys.75.1,DQD_RevModPhys.79.1217} 
The charge state of a single-level quantum dot can be labeled as either empty or occupied based on the absence or presence of the excess electron.
The system in the AA model consists of two coupled QDs, each connected to an electron reservoir maintained at a fixed chemical potential ($\mu_{L/R}$) and temperature $T$ (see \cite{AA_model}).
The energy level of each dot can be tuned to three possible values
$\epsilon_u,\ \epsilon_0,$ and $\epsilon_l$ with, $\epsilon_l < \epsilon_0 < \epsilon_u$.
Coulomb repulsion prevents the DQD from being occupied by more than a single excess electron.
Hence the possible \textit{occupation states} are (i) \textit{L}: the left dot contains the excess electron, (ii) \textit{R}: the right dot contains the excess electron, (iii) \textit{E}: both dots are unoccupied.
The charge state of the DQD is monitored continuously, and a feedback scheme is applied.
The electron reservoirs coupled to the left and right QD's are maintained at chemical potentials $\mu_L$ and $\mu_R$.
If $\mu_R> \mu_L$, then transferring an electron from the left to the right reservoir requires electrical work to be performed against a voltage bias.\par 
 
The protocol of the AA model \cite{AA_model} was originally introduced in Ref.\cite{Averin_PhysRevB.84.245448}. In the ideal mode of operation the DQD starts in the empty state, with the energy level of the left QD at $\epsilon_0$ and that of the right QD at  $\epsilon_u$, where $(\epsilon_u-\mu_{L/R})\gg k_BT$ and $k_B$ is Boltzmann's constant.
When an electron enters the left QD from the left reservoir, instant feedback is applied to change the energy levels of both the left and right QDs to $\epsilon_l$, where $(\mu_{L/R}-\epsilon_l)\gg k_BT$.
During this first feedback step, the external agent extracts $(\epsilon_0-\epsilon_l)$ work.
Next, the system is monitored until the electron tunnels from the left to the right QD, at which point feedback is
applied to change the energy level of the left QD to $\epsilon_u$
and the right QD to $\epsilon_0$. The external agent performs $(\epsilon_0 - \epsilon_l)$ work during this second feedback step, cancelling the work extraction of the previous step. Finally, when the electron jumps from the
right QD to the right reservoir, feedback is applied again to switch
the energy levels of the DQD back to their initial values. No work is performed during this step, as the DQD is empty. This three-step cyclic protocol transfers an electron from the left to the right reservoir.
Since no net work is performed by the external agent, the energy for this transfer must come from the thermal reservoirs.
Thus the feedback-driven cycle ultimately converts heat into chemical work, of the amount $W_{\text{ext}}=(\mu_R - \mu_L)$.
See Ref. \cite{AA_model} for more details.
The protocol discussed above is shown in Figure \ref{fig:DQD_states}, where electron transition events are indicated by solid arrows, and feedback steps by dotted arrows.

\begin{figure*}{
    \includegraphics[scale=0.6]{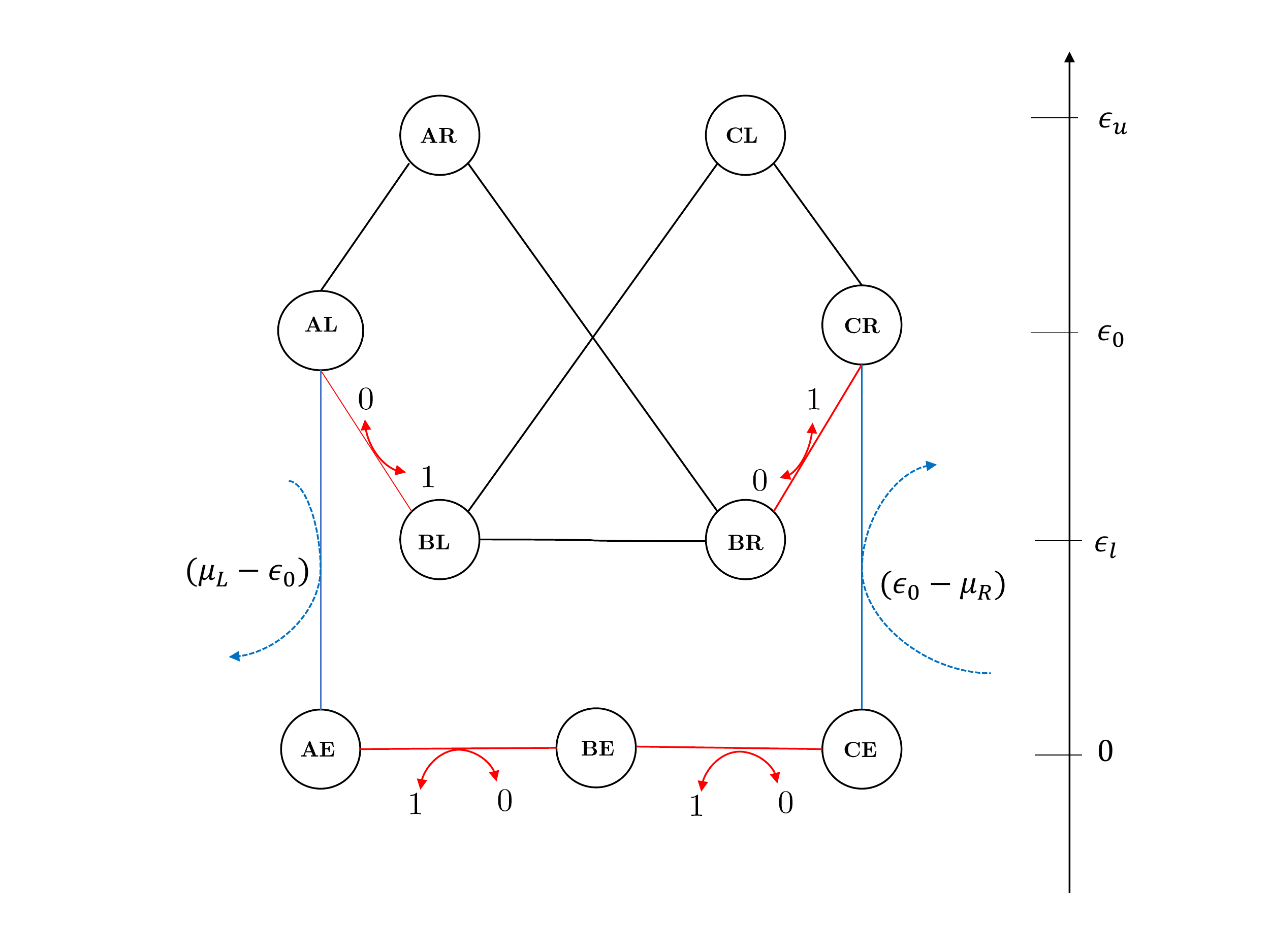}}
    \caption{Reduced network $\mathcal{G}_r=(\mathbf{V_x},\mathbf{E_x})$. Energies of the states in $\mathbf{V_x}$ are shown to the right of the network. Edges shown in red correspond to the feedback steps of the original AA model, and involve the flipping of the bit in the memory-tape model. The edges shown in blue correspond to the transitions where the electron hops into (out of) the DQD from (to) an electron reservoir and the dotted arrows show the corresponding energy exchange.
    \label{fig:DQD_network_1}}
\end{figure*}

\section{Memory-tape model of Maxwell's demon in DQD system}
\label{Sec:Stochastic_Modelling}
\subsection{Reduced Network: 9 states}
\label{Sec: 9_states}
\subsubsection{Network structure}
We now construct a network representation of the states of the AA model, as a first step toward designing a corresponding memory-tape model. In the AA model, the DQD occupation state $\sigma$ is a dynamic variable with three possible states, $\sigma \in\mathbf{\Sigma}=\{L, E, R\}$, as described above. The DQD \textit{energy configuration} $\lambda$ acts as a control parameter, also with three possible states: $\lambda \in \mathbf{\Lambda}=\{A,B,C\}$ where $A\equiv(\epsilon_0,\epsilon_u), \ B\equiv(\epsilon_l,\epsilon_l) \  \text{and }C\equiv(\epsilon_u, \epsilon_0)$. Combining the energy configurations and the occupation states leads to nine possible states for the \textit{DQD state variable}: $x \in \mathbf{V_x}=\mathbf{\Lambda} \times \mathbf{\Sigma}= \{AL,AE,\cdots CR\}$. The ideal cyclic protocol $\mathcal{C}_{AA}$, described above, follows the path $AE \longrightarrow AL \implies BL
\longrightarrow BR \implies CR \longrightarrow CE \implies AE$, where double arrows signify feedback steps (see Fig.~\ref{fig:DQD_states}).\par 

We now consider a situation in which the energy configuration $\lambda$ is no longer a control parameter, but instead is a dynamical variable on the same footing as the occupation state $\sigma$.
In our model, the entire system is maintained at a temperature $T$ using a thermal bath, and $\lambda$ is now a stochastic variable that evolves under the effect of the thermal noise from the bath.
The system-variable $x\equiv(\lambda,\sigma)$ evolves among the nine states in $\mathbf{V_x}$ under a Poisson jump process.
We make the following assumptions:
(i) The elementary transitions in our process involve a change in either $\lambda$, or $\sigma$, but not both simultaneously, i.e., the system is \textit{bipartite} \cite{Bipartite_2014}.
(ii) If $\lambda=B$, then the excess electron cannot hop in or out of the electron reservoirs; thus, the transitions $ BE  \leftrightarrow BL$ and $BE \leftrightarrow BR$ are not allowed.
(iii) Direct transitions between $A$ and $C$ states are forbidden.
These assumptions are modelling choices, but we note that all of the forbidden transitions can be justified physically by assuming sufficiently high energy barriers between corresponding states.\par 

Under these assumptions, we obtain a network $\mathcal{G}_r=(\mathbf{V_x},\mathbf{E_x})$ where $\mathbf{V_x}\equiv V(\mathcal{G}_r)$ is the set of 9 vertices and $\mathbf{E_x}\equiv E(\mathcal{G}_r)$ is the set of 11 bidirectional edges (see Fig.~\ref{fig:DQD_network_1}), describing the stochastic dynamics \cite{Schnakenberg_RevModPhys_1976,Seifert_FT_review_2012,Bipartite_2014} of the variable $x\equiv(\lambda,\sigma)$.
The subscript $r$ in $\mathcal{G}_r$ indicates a \textit{reduced} 9-state network, in contrast with a \textit{full} 18-state network $\mathcal{G}_f$ to be defined later.

\subsubsection{Dynamics in the reduced network}

We set the energies of the empty states $AE,BE,CE$ to zero and assign energies to all other states based on the energy level of the QD that contains the electron: states $BL$ and $BR$ have energy $\epsilon_l$; states $AL$ and $CR$ have energy $\epsilon_0$;
  and states $AR$ and $CL$ have energy $\epsilon_u$, with $\epsilon_l < \epsilon_0 < \epsilon_u$ as mentioned earlier. We impose the condition of local detailed balance on the transition rates for the thermal transitions $x_i \leftrightarrow x_j$ with $x_i,x_j \in V(\mathcal{G}_r)$, when there is no exchange of electron with the left or the right reservoir:
 \begin{equation}\label{Detailed_balance_eqn_reduced}
    \frac{R^{r}_{x_ix_j}}{R^{r}_{x_jx_i}}=e^{-\beta(E^{r}_i -E^{r}_j)} \quad ,
\end{equation}
where $\beta = (k_BT)^{-1}$ is the inverse temperature, and the superscript $r$ again refers to the reduced network. $E^{r}_i$ ($E^{r}_j$) is the energy of the state $x_i$ ($x_j$) and $R^{r}_{x_ix_j}$ is the transition rate for the jump $x_j  \rightarrow  x_i $.Strictly speaking, the tunneling events of the excess electron between two QDs (i.e., $\sigma= L \ \leftrightarrow \ \sigma =R$) are coherent transfers, a purely quantum phenomenon and not classical thermal jumps in the original AA model as discussed in Sec.~\ref{Sec: background_setup}. However, for simplicity, we model these transitions as classical thermal jumps and assume the local detailed balance relation Eq.~\ref{Detailed_balance_eqn_reduced} for the edges: $AL\leftrightarrow AR,\  BL \leftrightarrow BR\text{ and }  CL \leftrightarrow CR $.

When an electron jumps from the right reservoir, maintained at the chemical potential $\mu_R$ to the energy level $\epsilon_0$ of
the right QD, there is an energy cost of $(\epsilon_0 - \mu_R)$ and similarly if an electron jumps from the level $\epsilon_0$ of the left QD to the left electron reservoir set at the chemical potential $\mu_L$ the energy exchange is $(\mu_L -\epsilon_0$). Thus for the transitions $AL \leftrightarrow AE$ and $CR \leftrightarrow CE$ (shown in blue in Fig.~\ref{fig:DQD_network_1}), we can write the local detailed balance relations as,
\begin{equation}\label{reservoir_eqns_reduced}
\begin{split}
     \frac{R^{r}_{AE \ AL}}{R^{r}_{AL \ AE}} =  e^{-\beta(\mu_L - \epsilon_0)}, \\
     \frac{R^{r}_{CR \ CE}}{R^{r}_{CE \ CR}}= e^{-\beta(\epsilon_0 - \mu_R)}\text{.}
\end{split}
\end{equation}
The coupling with the electron reservoir creates \textit{thermodynamic forces} \cite{Seifert_FT_review_2012,Schnakenberg_RevModPhys_1976,Hill_Simmons_FreeEnergyTransduction_76,hill2013free} in $\mathcal{G}_r$ and leads to violation of global detailed balance when $\mu_L \ne \mu_R$.  
When Eqs.~\ref{Detailed_balance_eqn_reduced} and \ref{reservoir_eqns_reduced} are satisfied and $\mu_L \ne \mu_R$, the dynamics of $x$ in $\mathcal{G}_r$ reach a non-equilibrium steady state (NESS) \cite{Seifert_FT_review_2012}.
In this state, electrons flow in the thermodynamically preferred direction, i.e., from the right (left) reservoir to the left (right) reservoir when $\mu_R> \mu_L$ ($\mu_L>\mu_R$), resulting in an overall counterclockwise (clockwise) flow of probability current in $\mathcal{G}_r$.
This flow is in contrast with the feedback-controlled model, which transfers electrons against the thermodynamically preferred direction.
Therefore we next consider how to couple the DQD to an information reservoir, in the form of a stream of bits, so as to make the evolution of the DQD mimic that of the AA model.

\subsection{Bit coupling strategy}
\label{Sec:bit_coupling}
Our information reservoir is a memory-tape containing $n$ classical bits.
Each bit ($b$) can be in one of the two states in $\mathbf{B}=\{0,1\}$.
The energies of the two bit states are degenerate, and we set them to zero.
As in Ref.~\cite{MJ_model} the DQD interacts with a bit for an interval of duration $\tau$, after which the next bit arrives.
We can visualize this process by imagining that the bits are placed, equally spaced, on a tape that moves frictionlessly past the DQD, which interacts with the bit that is nearest to it at any given time.

In our model the coupling between the DQD and the bit occurs along the four edges of $\mathcal{G}_r$ that correspond to instant feedback steps in the AA model. These edges are shown in red in Fig.~\ref{fig:DQD_network_1}. (Note that we have split the $CE \implies AE$ feedback step of the original AA model into two steps: $CE\leftrightarrow BE \text{ and } BE\leftrightarrow AE$ in our model.)
Specifically, the DQD transitions corresponding to these four edges can  occur only when the state of the interacting bit $b$ also flips.
We set up the coupling rules so that the CW flow of probability current along $\mathcal{C}_{AA}$ is favoured when $b$ flips from $0$ to $1$, and CCW flow is favored when $b$ flips from $1$ to $0$. 
For example, the transition $AL  \rightarrow  BL$ must be accompanied by a bit flip $0\rightarrow 1$, and the reverse transition $BL  \rightarrow  AL$ occurs only if the interacting bit flips from $1$ to $0$.
Similar comments apply to the edges $BR \leftrightarrow CR$, $CE \leftrightarrow BE$ and $BE \leftrightarrow AE$.
These DQD-bit coupling rules are indicated by curved red arrow in Fig.~\ref{fig:DQD_network_1}.
 With this coupling scheme, an excess of $0$'s in the incoming bit stream biases the flow of probability in the CW direction.
This bias opposes the thermodynamic direction of the probability current when $\mu_R> \mu_L$.
Similarly, if $\mu_L> \mu_R$ then an excess of $1$'s opposes the thermodynamic direction of the probability current.\par
 
\subsection{Full Network: 18 states}
\label{Sec: 18states}
\subsubsection{Network structure}
\begin{figure*}{
    \includegraphics[scale=0.6]{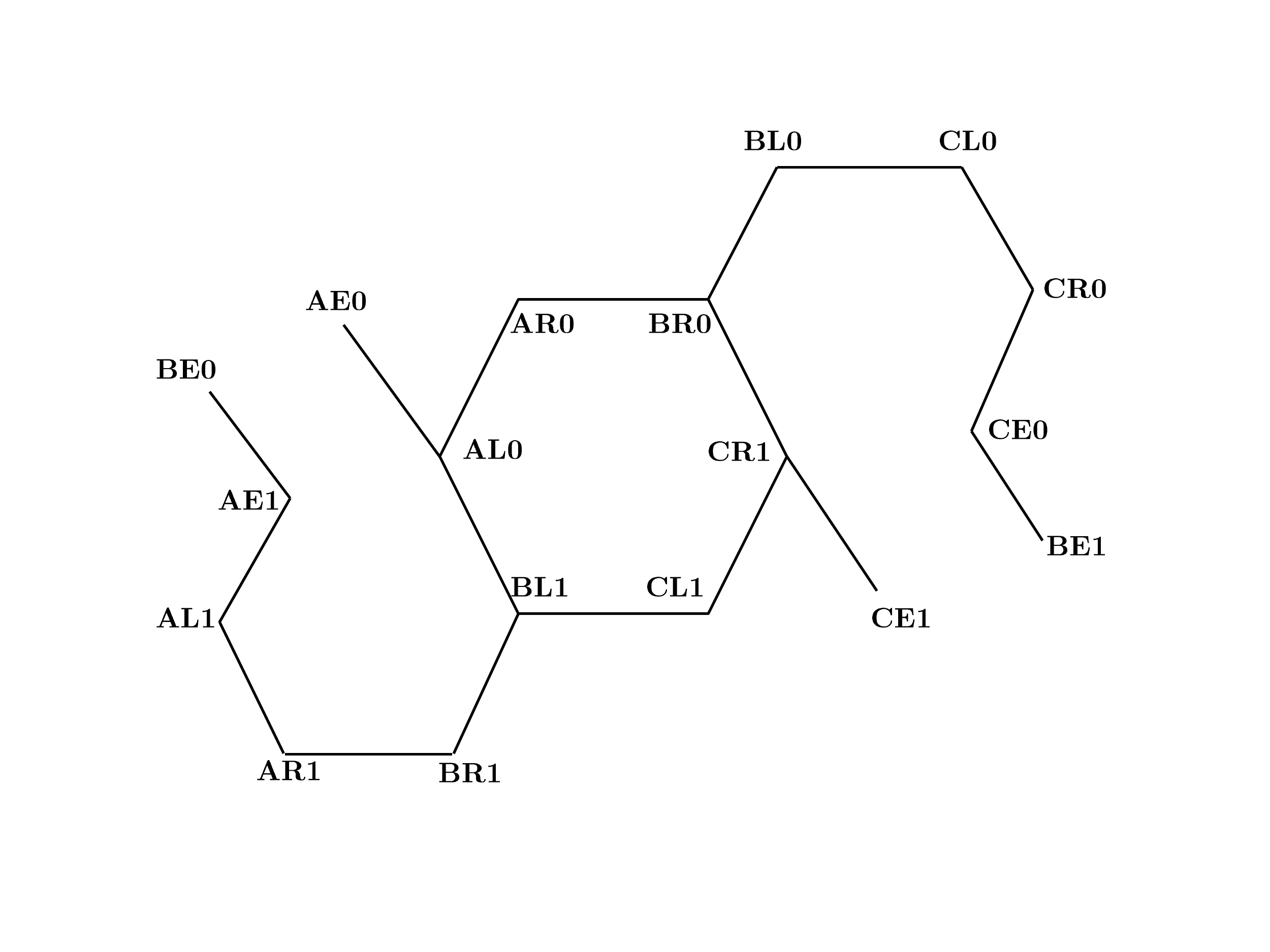}}
    \caption{Full network $\mathcal{G}_f=(\mathbf{V_y},\mathbf{E_y})$ showing all 18 states of the combined DQD and bit. Eq.~\ref{Master_eqn} governs the dynamics of the variable $y\equiv(\lambda,\sigma,b)$ in this network.}
    \label{fig:DQD_network_2}
\end{figure*}
The joint evolution of the DQD state ($x$) and the nearest bit ($b$) occurs in the \textit{full network} $\mathcal{G}_f=(\mathbf{V_y},\mathbf{E_y})$ (see Fig.~\ref{fig:DQD_network_2}). 
Here $\mathbf{V_y}\equiv V(\mathcal{G}_f) =\mathbf{V_x}\times \mathbf{B}$ is the set of vertices representing the 18 possible states of the variable $y=(x,b) \equiv(\lambda,\sigma,b)$, and $\mathbf{E_y}\equiv E(\mathcal{G}_f)$ is the set of 18 bidirectional edges that are drawn based on the bit-coupling rules described in Sec.~\ref{Sec:bit_coupling}.
Each edge of $\mathcal{G}_r$ that does not involve bit coupling is represented by two different edges of $\mathcal{G}_f$, corresponding to the two possible bit states.
That is, an edge $x_i \leftrightarrow x_j\in E(\mathcal{G}_r)$ corresponds to the edges $(x_i,0) \leftrightarrow (x_j,0)$ and $(x_i,1) \leftrightarrow (x_j,1)$ in $E(\mathcal{G}_f)$, when $x_i \leftrightarrow x_j$ does not involve bit coupling.
An edge $x'_i \leftrightarrow x'_j \in E(\mathcal{G}_r)$ that is coupled to the bit transition $0 \leftrightarrow 1$ is mapped to only one edge $(x'_i,0) \leftrightarrow (x'_j,1) \in E(\mathcal{G}_f)$. There are four such edges in $\mathcal{G}_f$: $BE0 \leftrightarrow AE1, \  AL0  \leftrightarrow BL1, \  BR0  \leftrightarrow  CR1, \ \text{ and } CE0 \leftrightarrow BE1$ (see Figs.~\ref{fig:DQD_network_1} and \ref{fig:DQD_network_2}).  \par 

\subsubsection{Dynamics in the full network}
\label{Sec:Energetics}

As the $b=0$ and $1$ bit states are energetically degenerate, the transition rates for the edges in $E(\mathcal{G}_f)$ obey the same detailed balance conditions as the corresponding edges in $E(\mathcal{G}_r)$.
Edges $y_i\leftrightarrow y_j$ in $E(\mathcal{G}_f)$ with no electron reservoir coupling satisfy
\begin{equation}\label{Detailed_balance_eqn}
    \frac{R_{y_iy_j}}{R_{y_jy_i}}=e^{-\beta(E_i -E_j)},
\end{equation}
where $E_i$ and $E_j$ are the energies of the states $y_i$ and $y_j$ respectively (compare Eq~\ref{Detailed_balance_eqn} with Eq.~\ref{Detailed_balance_eqn_reduced}). When there is a coupling with the electron reservoirs, the local detailed balance relations are given as,
\begin{equation}\label{reservoir_eqns}
\begin{split}
     \frac{R_{AE0 \ AL0}}{R_{AL0 \ AE0}}= \frac{R_{AE1 \ AL1}}{R_{AL1 \ AE1}}=e^{-\beta(\mu_L - \epsilon_0)},\\
     \frac{R_{CR0 \ CE0}}{R_{CE0 \ CR0}}=\frac{R_{CR1 \ CE1}}{R_{CE1 \ CR1}}=e^{-\beta(\epsilon_0 - \mu_R)}\text{.}
\end{split}
\end{equation}
(compare Eq.~\ref{reservoir_eqns} with Eq.~\ref{reservoir_eqns_reduced}).
Eqs.~\ref{Detailed_balance_eqn} and \ref{reservoir_eqns} ensure the thermodynamic consistency of the model, but do not yet completely specify the dynamics of $y$. We assume the slowest timescale of the stochastic dynamics of $y$ is on the order of unity, and our choice of the transition rates consistent with Eqs.~\ref{Detailed_balance_eqn} and \ref{reservoir_eqns} are shown in Table~\ref{Ratematrixtable}.\par 

During every interaction interval of duration $\tau$, the joint dynamics of the DQD and bit are described by a Poisson jump process for the state variable $y=(x,b)$ in $\mathcal{G}_f$, with transition rates shown in Table~\ref{Ratematrixtable}. At the end of each interaction interval, when a new bit $b_{in}$ arrives, the state of the DQD $x$ remains unchanged, and the state of the interacting bit $b$ takes on the value of the incoming bit $b_{in}$. Thus when the outgoing and incoming bit states differ, there is an effective \textit{virtual jump}, due to the fact that the ``old'' interacting bit is replaced by the next bit in the memory tape. 

\subsection{Summary of the modelling strategy}
\label{Sec:Summary_modelling_strategy}

Here we summarize our approach for creating an autonomous, memory-tape model of Maxwell's demon from the non-autonomous, feedback-controlled AA model. We first create a network representation of the states of the feedback-controlled model by identifying the dynamical states of the system $(\sigma \in \mathbf{\Sigma})$ and the states of the control parameter$(\lambda \in \mathbf{\Lambda})$. We then convert the control parameter $\lambda$ to a stochastic dynamic variable that jumps among the states of $\mathbf{\Lambda}$. The joint state of the system and parameter is given by $x\equiv(\lambda,\sigma) \in \mathbf{V_x}$.
The next step is to identify a network $\mathcal{G}_r=(\mathbf{V_x},\mathbf{E_x})$ whose edges correspond to possible transitions. For thermodynamic consistency, the transition rates must satisfy Eqs.~\ref{Detailed_balance_eqn_reduced} and \ref{reservoir_eqns_reduced}.  There is no unique way to construct the network $\mathcal{G}_r $ and different choices of the allowed transitions lead to different memory-tape models. For our DQD system, we focused on designing a model that mimics the feedback-controlled model's behavior, and is simple enough for analytical and semi-analytical treatment.

Next, the DQD is connected to a sliding memory-tape (information reservoir). By interacting with only the nearest bit on the tape, the DQD interacts with each bit for a fixed time $\tau$. During that time, the coupling between the DQD and the interacting bit $b$ occurs along those edges in the network $\mathcal{G}_r$ that  correspond to the instantaneous feedback steps of the AA model. 
The bit coupling rules are chosen so that incoming bits in the $0$ state bias the resulting current in one direction (CW in our model) and incoming bits in the $1$ state bias it in the other direction.  In this way a memory tape with a surplus of $0$'s or $1$'s generates an effective force, which can be harnessed to oppose the thermodynamic forces arising from (for example) reservoirs at different chemical potentials.

The joint state of the DQD and interacting bit is described by a variable $y\equiv(x,b)$ that evolves by a Poisson jump process in the network $\mathcal{G}_f=(\mathbf{V_y,E_y})$. As we assume the bit states $0$ and $1$ to be energetically degenerate, the transition rates in the $\mathcal{G}_f$ follow from those in $\mathcal{G}_r$ (see Eqs.~\ref{Detailed_balance_eqn} and \ref{reservoir_eqns}).\par

\begin{table}

\caption{\label{Ratematrixtable} Transition rates for jumps of the variable $y$ in $\mathcal{G}_f$. $R_{y_iy_j}$ denotes the transition rate from $y_j$ to  $y_i$. Here we have taken $r=e^{-\beta \epsilon}$ with $\epsilon= (\epsilon_u -\epsilon_0)= (\epsilon_0-\epsilon_l)$. These rates are used construct the matrix $\mathbf{R}$ which is shown in Eq.~\ref{eq:rate_matrix_full} in Appendix~\ref{app:RateMatrix}}.
\begin{ruledtabular}
\begin{tabular}{c}
$R_{CL0\ CR0}=r$ \\
$R_{CR0\ CL0}=1$ \\
$R_{CL1\ CR1}=r$\\
$R_{CR1\ CL1}=1$\\
$R_{BL0\ CL0}=1$\\
$R_{CL0\ BL0}=r^2$\\
$R_{BL1\ CL1}=1$\\
$R_{CL1\ BL1}=r^2$\\
$R_{BL0\ BR0}=1$\\
$R_{BR0\ BL0}=1$\\
$R_{BL1\ BR1}=1$\\
$R_{BR1\ BL1}=1$\\
$R_{AR0\ BR0}=r^2$\\
$R_{BR0\ AR0}=1$\\
$R_{AR1\ BR1}=r^2$\\
$R_{BR1\ AR1}=1$\\
$R_{AL0\ AR0}=1$\\
$R_{AR0\ AL0}=r$\\
$R_{AL1\ AR1}=1$\\
$R_{AR1\ AL1}=r$\\
$R_{AE0\ AL0}=e^{-\beta(\mu_L-\epsilon_0)}$\\
$R_{AL0\ AE0}=1$\\
$R_{AE1\ AL1}=e^{-\beta(\mu_L-\epsilon_0)}$\\
$R_{AL1\ AE1}=1$\\
$R_{CR0\ CE0}=1$\\
$R_{CE0\ CR0}=e^{-\beta(\mu_R-\epsilon_0)}$\\
$R_{CR1\ CE1}=1$\\
$R_{CE1\ CR1}=e^{-\beta(\mu_R-\epsilon_0)}$\\
$R_{BE0\ AE1}=1$\\
$R_{AE1\ BE0}=1$\\
$R_{CE0\ BE1}=1$\\
$R_{BE1\ CE0}=1$\\
$R_{AL0\ BL1}=r$\\
$R_{BL1\ AL0}=1$\\
$R_{BR0\ CR1}=1$\\
$R_{CR1\ BR0}=r$\\
\end{tabular}
\end{ruledtabular}
\end{table}

\section{Analysis and Results}
\label{Sec:Analyses_Results}

\subsection{Methods}
\label{Sec:Analyses_methods}
Following Ref.~\cite{MJ_model}, let $\mathbf{p}(t_n)$ be a column vector with nine entries that describes the probability distribution of the states of the DQD state variable $x$ in $\mathcal{G}_r$ (in the order $AE$, $BE$, $CE$, $BL$, $BR$, $AL$, $CR$, $AR$, $CL$) at time $t_n \equiv n\tau$ that marks the start of an interaction interval.
Each incoming bit is independently sampled from the same probability distribution, with $p_0$ (or $p_1$) denoting the probability of the bit to arrive in state $0$ (or $1$). It is convenient to specify this distribution by the single parameter $\delta=p_0-p_1$, which measures the excess of $0$'s among the incoming bits.
The statistical state of the variable $y\equiv(x,b)$ in $\mathcal{G}_f$ at time $t_n$ (just after the arrival of the $n$'th bit) is given by the 18-dimensional vector 
\begin{equation}
\mathbf{p_f}(t_n)=\mathbf{Mp}(t_n), \ \
\mathbf{M}= \begin{pmatrix}
p_0 \mathbf{I} \\
p_1 \mathbf{I}
\end{pmatrix},
\end{equation}
with $\mathbf{I}$ being a $9 \times 9$ identity matrix. The first nine elements of $\mathbf{p_f}(t)$ correspond to the bit state $b=0$ and the last nine elements to the the state $b=1$.
From $t=t_n$ to $t_{n+1}$ the probability distribution in $\mathcal{G}_f$ evolves under the master equation
\begin{equation}\label{Master_eqn}
    \dv{}{t}\mathbf{p_f}(t) =\mathbf{R  p_f}(t),
\end{equation}
where $\mathbf{R}$ is the $18\times18$ rate matrix whose off-diagonal elements are the transition rates listed in Table~\ref{Ratematrixtable}, and whose diagonal elements are $R_{y_iy_i}=-\sum_{y_j \ne y_i}R_{y_jy_i}$ (see Eq.~\ref{eq:rate_matrix_full} for an explicit expression for $\mathbf{R}$). 
At the end of the interaction interval, just before the next bit arrives, the joint probability distribution is obtained from the solution of Eq.~\ref{Master_eqn}, namely
\begin{equation}
    \mathbf{p_f}(t_n+\tau)=e^{\mathbf{R}\tau}\mathbf{Mp}(t_n).
\end{equation}\par

To obtain the probability distribution of $x$ in $\mathcal{G}_r$ at the end of the interaction interval, we project from the 18-state network $\mathcal{G}_f$ to the 9-state network $\mathcal{G}_r$,
\begin{equation}\label{PD_eqn_time_evolve}
    \mathbf{p}(t_n+\tau)=\mathbf{P_D } e^{\mathbf{R}\tau} \mathbf{M} \mathbf{p}(t_n),\ \ \mathbf{P_D}=\begin{pmatrix}
    \mathbf{I} & \mathbf{I}
    \end{pmatrix}.
\end{equation}
Equivalently,
\begin{equation}\label{Transition_matrix_map}
   \mathbf{p}((n+1)\tau)=\mathbf{T}\mathbf{p}(n\tau), \ \  \mathbf{T}=\mathbf{P_D }e^{\mathbf{R}\tau}\mathbf{M}.
\end{equation}
This transition matrix $\mathbf{T}$ (which depends on the parameter $\tau$) evolves the probability distribution of $x$ in $\mathcal{G}_r$ over a single interaction interval.
The evolution over $n$ successive intervals is described by the transition matrix $\mathbf{T}^n$.
From the Perron-Frobenius theorem \cite{PerronFrobTheorem_meyer2000matrix} it follows that any distribution $\mathbf{p}$ in $\mathcal{G}_r$ evolves asymptotically to a unique periodic steady state
\begin{equation}
    \mathbf{q_{\text{pss}}}= \lim_{n \to \infty} \mathbf{T}^n \mathbf{p} \quad .
\end{equation}
The periodic steady state $\mathbf{q_{\text{pss}}}$ is a function of the interaction interval $\tau$, and can be calculated by solving for the invariant vector of the matrix $\mathbf{T}$, 
\begin{equation}\label{steady_state_eqn}
    \mathbf{T} \ \mathbf{q_{\text{pss}} }= \mathbf{q_{\text{pss}}}.
\end{equation}

Once the periodic steady state for the DQD has been reached, the joint state of the DQD and bit at the start of every interaction interval is given by $\mathbf{M} \mathbf{q_{\text{pss}}}$,
and the joint state at a time $t_n+\Delta t$, with $0\le\Delta t<\tau$, is
\begin{equation}
\label{eq:pftn+tau}
    \mathbf{p_f}(t_n+\Delta t) = e^{\mathbf{R}\Delta t} \mathbf{M} \mathbf{q_{\text{pss}}} \quad .
\end{equation}
For the remainder of this paper, when analyzing the behavior of our model, we will assume that the periodic steady state has been reached.

\subsection{Thermodynamics of the memory-tape model}
\label{Sec:Thermo}

\subsubsection{Calculation of work}

Let the \textit{circulation} $\Phi(\tau)$ denote the average number of electrons transferred from the left to the right reservoir during each interaction interval.
The average chemical work performed by the DQD system per time interval $\tau$ is then
\begin{equation}
    \label{work_def}
    W(\tau)=(\mu_R - \mu_L) \Phi(\tau). 
\end{equation}
If the sign of $\mu_R-\mu_L$ is the same as that of $\Phi(\tau)$, then electrons flow from the lower to higher chemical potential, that is against the thermodynamic force.

From Fig.~\ref{fig:DQD_network_1} we see that
\begin{equation}
\label{eq:circulation_def}
\begin{split}
    \Phi(\tau)=\int_0^\tau \ dt \ J^r_{CR \to CE}  =\int_0^\tau  \ dt \ J^r_{CE \to BE} \\ =\int_0^\tau  \ dt \ J^r_{BE \to AE} =\int_0^\tau  \ dt \ J^r_{AE \to AL},
\end{split}
\end{equation}
where $J^r_{x_j \to x_i} \equiv J^r_{x_ix_j}$ is the probability current along $x_j\rightarrow x_i$ in $\mathcal{G}_r$, projected from the corresponding currents in $\mathcal{G}_f$.
We can determine $\Phi(\tau)$ by calculating any one of these integrals.

The probability current along $y_j\rightarrow y_i$ of $\mathcal{G}_f$ is 
\begin{equation}
    \label{current_def_full}
    J_{y_iy_j}=R_{y_iy_j}p_{y_j}(t)-R_{y_jy_i}p_{y_i}(t)\text{.}
\end{equation}
When two edges $x_j0\leftrightarrow x_i0$ and $x_j1 \leftrightarrow x_i1$ in $\mathcal{G}_f$ correspond to the edge $x_j \leftrightarrow x_i$ in $\mathcal{G}_r$, we have
\begin{equation}
   J^r_{x_ix_j}(t)=J_{x_i0 \ x_j0}(t) + J_{x_i1 \ x_j1}(t), 
\end{equation}
but when the transition $x_j \rightarrow x_i$ is coupled with a bit flip $b'  \rightarrow b''$, we have
\begin{equation}
\label{eq:proj_curr_bit_flip}
   J^r_{x_ix_j}(t)=J_{x_ib'' \ x_jb'}(t)\text{.}
\end{equation}
 \par 

Since the $CE \leftrightarrow BE$ transition is coupled to the bit flit $0 \leftrightarrow 1$, the edge $CE \leftrightarrow BE$ in $\mathcal{G}_r$ corresponds to a single edge,  $CE0\leftrightarrow BE1$ in $\mathcal{G}_f$, hence
\begin{equation}
    \Phi(\tau) = \int_{0}^{\tau} \dd{t} \ J^r_{BE \ CE}= \int_{0}^{\tau} \dd{t} \ J_{BE1 \ CE0}.
\end{equation}
Moreover, since $BE1$ is connected to only one edge, $CE0\leftrightarrow BE1$, we have $\dot{p}_{BE1}=J_{BE1 \ CE0}$.
Therefore,
\begin{eqnarray}
\label{circulation}
    \Phi(\tau) = \int_{0}^{\tau} \dd{t} \ \dot{p}_{BE1} &=& [p_{BE1}(\tau) - p_{BE1}(0)] \nonumber \\
    &=& \left[ \left( e^{\mathbf{R}\tau} - \mathbf{I} \right) \mathbf{M q_{\text{pss}}}\right]_{y=BE1} \quad ,
\end{eqnarray}
where we have used Eq.~\ref{eq:pftn+tau} to get to the second line.
We will use this result in Sec.~\ref{SubSec:Analytical_results-therm}.

\subsubsection{Calculation of entropy change of the bit}
Let $p_0'$ and $p_1'$ denote the probabilities of the outgoing bit to be in the states $0$ and $1$.
These values are determined by summing over the appropriate states $y=(x,b)$ in $\mathcal{G}_f$ at the end of an interaction interval:
\begin{equation}
\label{final_bit_dist}
\begin{split}
    p_0'&=\sum_{x\in V(\mathcal{G}_r)} (e^{\mathbf{R}\tau}\mathbf{M \ q_{\text{pss}}})_{y=(x,0)} \\ 
    p_1'&=\sum_{x\in V(\mathcal{G}_r)} (e^{\mathbf{R}\tau}\mathbf{M \ q_{\text{pss}}})_{y=(x,1)} \quad .
\end{split}
\end{equation}
The parameter
\begin{equation}
\label{delta_out}
    \delta'=p'_0-p'_1,
\end{equation}
specifies the distribution of the outgoing bit. The entropy corresponding to this distribution is $S'=-\sum_{i=0,1}p_i'\ln{p_i'}$, while that of the incoming bit is $S=-\sum_{i=0,1}p_i\ln{p_i}$. Thus in the periodic steady state, the change in \textit{single-symbol entropy} \cite{Boyd2016} of the interacting bit is $\Delta S=S'-S$. 
Because $\Delta S$ does not account for correlations that develop between successive outgoing bits, it provides only an upper bound on the net entropy change (per bit) of the information reservoir.
We discuss this point in detail in the next section (\ref{subsec: 1st_2nd_law_general}), in the context of the second law of thermodynamics.

\subsubsection{The first and the second law of thermodynamics}
\label{subsec: 1st_2nd_law_general}
In the periodic steady-state, the change in the internal energy of the DQD over one interaction interval is zero, on average. If chemical work is performed by the flow of electrons from low to high chemical potential, then the energy required for this process must be extracted as heat from the thermal reservoir that maintains the entire system at a fixed temperature $T$.
We write the first law of thermodynamics at the periodic steady state for this model as
\begin{equation}
    \label{1st_law}
    Q(\tau)=W(\tau)=(\mu_R - \mu_L)\Phi(\tau),
\end{equation}
where $Q(\tau)$ is the average heat extracted from the thermal reservoir, per interaction interval.\par

In Refs. \cite{Boyd2016,boyd2017leveraging}, a general form of the second law for the information ratchets, called the \textit{Information Processing Second Law} (IPSL), was derived. In the periodic steady state the IPSL is written as
\begin{equation}
    \label{IPSL}
 ( \ln{2}) \Delta h_\mu\ge \beta W, 
\end{equation}
where $\Delta h_\mu$ is the change in the \textit{Shannon entropy rate} (see Ref. \cite{Boyd2016,boyd2017leveraging}) and $W$ is the average work extracted in one interaction interval. The entropy rate $\Delta h_\mu$  includes the effect of correlations among the bits in the incoming and outgoing bit-streams. In our model we have assumed that incoming bits are uncorrelated with each other and have been generated through a \textit{memoryless} \cite{boyd2017leveraging} process. For finite $\tau$, the outgoing bits become correlated with each other, and thus the output is \textit{memoryful} \cite{boyd2017leveraging}.
However, in the limit $\tau\rightarrow\infty$ these correlations become lost, and the Shannon entropy rate $\Delta h_\mu$ reduces to the change in single-symbol entropy $\Delta S /( \ln{2})$, hence for our model Eq.~\ref{IPSL} becomes (in that limit)
\begin{equation}
\label{second_law}
    \Delta S \ge \beta W \quad .
\end{equation}
Eq.~\ref{IPSL} is a general result for any memory-tape Maxwell's demon and Eq.~\ref{second_law} is a limiting case of the IPSL when correlations are neglected. When correlations are non-negligible, Eq.~\ref{IPSL} can identify functional modes of operation that are not indicated by Eq.~\ref{second_law} (see Ref. \cite{Boyd2016,Boyd_PRE_2017_correlation_powered_demon,Jurgens_PRR_2020}). However, it is common to use the single symbol entropy for the analysis of memory-tape models \cite{MJ_model,MJ_refrigarator,Barato_Seifert_EPL_2013,Barato_Seifert_PRL_2014,Strasberg_PRE_2014_spin_valve} and Eq.~\ref{second_law} has been previously derived in the context of Hamiltonian dynamics \cite{Deffner_CJ_PRX_2013} and stochastic dynamics \cite{Barato_Seifert_PRE_2014}. In our model, we ignore the effect of the correlations among the bits for simplicity and assume the validity of Eq.~\ref{second_law} as an approximation to Eq.~\ref{IPSL} even for finite $\tau$. The analysis of the effect of correlations among the bits and calculation of $\Delta h_\mu$ is outside the scope of this article (see Ref. \cite{Jurgens_PRR_2020} for calculation of $\Delta h_\mu$ in context of the MJ model). Henceforth, by ``entropy'' we always refer to single-symbol entropy unless otherwise specified.

\begin{figure*}[t!]
   \begin{subfigure}[$\delta$ variation]{
   \label{fig:delS_and_W_compare_long_time_a}
    \includegraphics[scale=0.25]{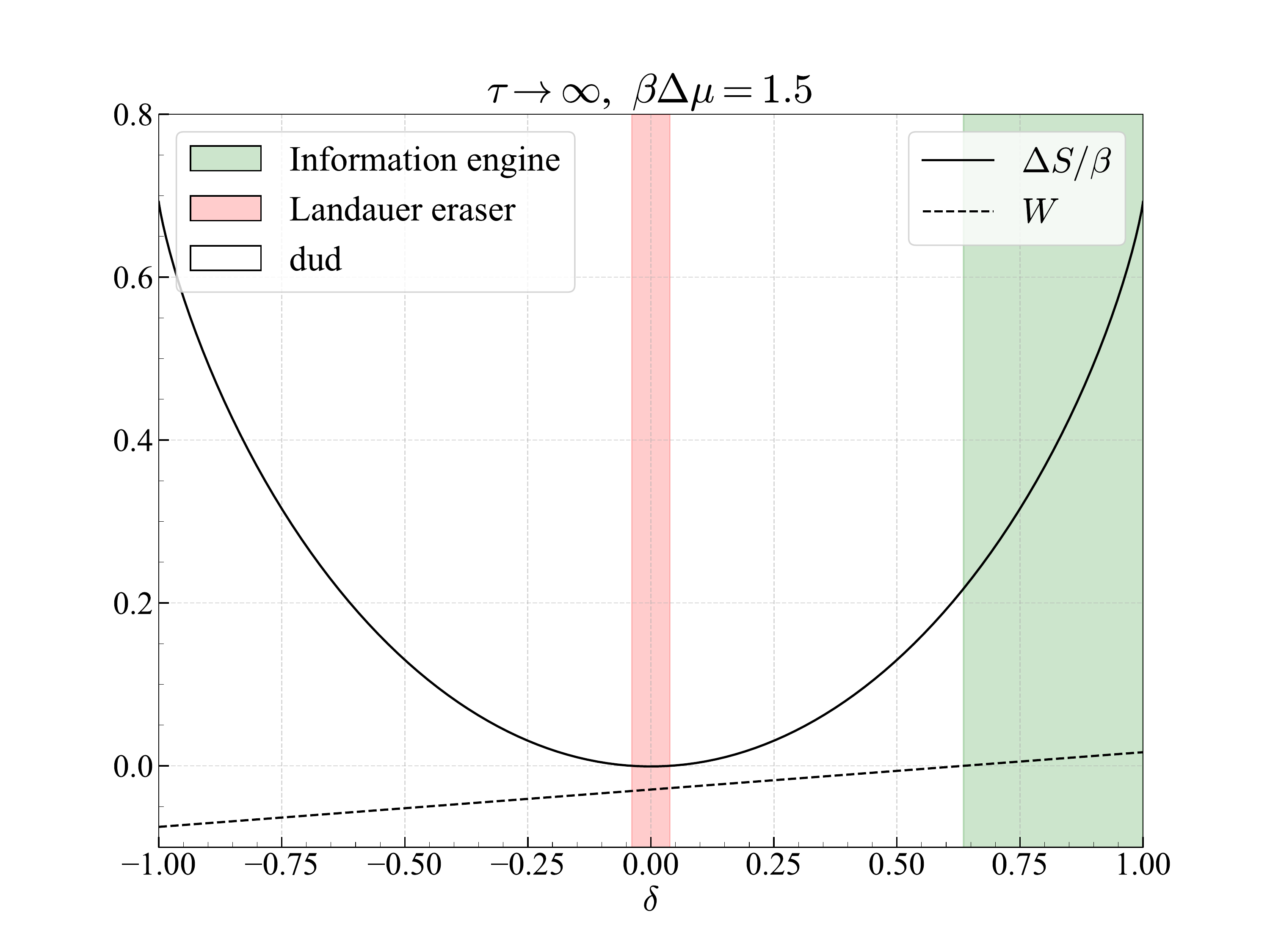}}
   \end{subfigure}
   \begin{subfigure}[$\Delta \mu$ variation]{
   \label{fig:delS_and_W_compare_long_time_b}
    \includegraphics[scale=0.25]{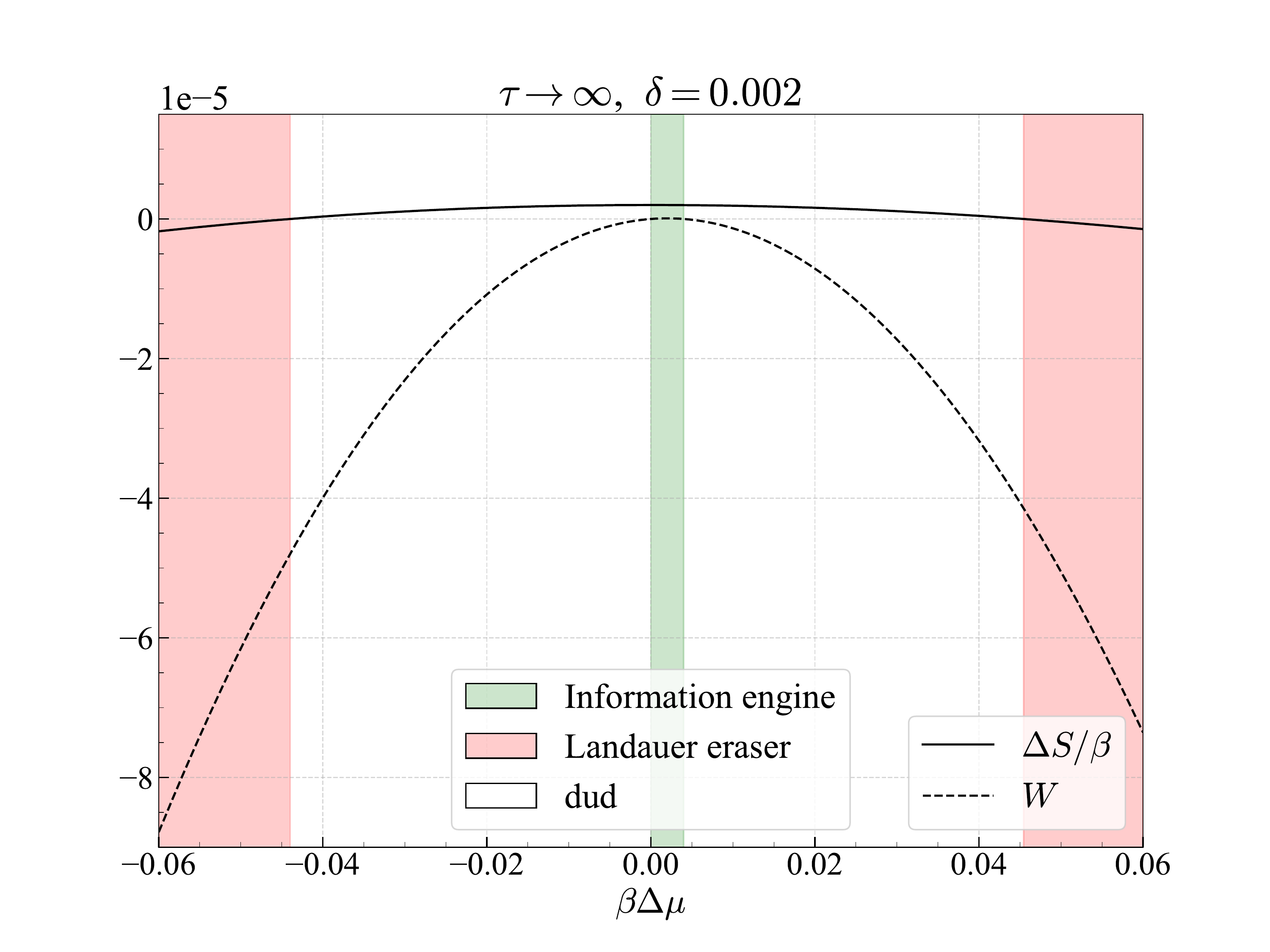}}
   \end{subfigure}
    \caption{Plots of $\Delta S/\beta$ and $ W$ when (a) $\delta$ is varied at fixed $\mu_R=1.5$ and $\mu_L =0$, and (b) $\beta \Delta \mu$ is varied by changing $\mu_R$ at fixed $\mu_L =0$ and $\delta=0.002$.  In both cases we set $\beta =1,\ r=e^{-1}, \ \epsilon_0 =0$, and we take the limit $\tau\rightarrow\infty$.
    In both plots we see that $\Delta S \ge \beta W$ is satisfied. The regions corresponding to the information engine $(\Delta S >0, W> 0)$, Landauer eraser $(\Delta S<0, W<0)$ and dud $(\Delta S> 0, W<0)$  are shaded green, red and white respectively.}
    \label{fig:delS_and_W_compare_long_time}
\end{figure*}

\subsection{Analytical results for $\tau \to \infty$}
\label{Sec:Analytical_results}
  
\subsubsection{Thermodynamic quantities}
\label{SubSec:Analytical_results-therm}
There are two relevant time scales in our model. We have taken the time scale associated with the thermal jumps in $\mathcal{G}_f$, which are governed by Eq.~\ref{Master_eqn}, to be of order unity. The other time scale is the parameter $\tau$ that defines how long the DQD interacts with each bit. From the Perron-Frobenius theorem \cite{PerronFrobTheorem_meyer2000matrix}, we have
\begin{equation}
    \label{stationary_state_full}
    \lim_{\tau \to \infty}e^{\mathbf{R}\tau}\mathbf{p_f}=\mathbf{\Pi}, \ \ \forall  \ \mathbf{p_f},
\end{equation}
where $\mathbf{R\Pi}=\mathbf{0}$. The expression for $\mathbf{\Pi}$ is given in Eq.~\ref{eq:stationary_dist_full_app} in Appendix~\ref{app:RateMatrix}. If $\tau$ is sufficiently large then Eq.~\ref{steady_state_eqn} becomes,
\begin{equation}
    \label{stationary_longtime}
     \mathbf{q}_{\text{pss}}^{\infty}=\lim_{\tau \to \infty}\mathbf{P_D}e^{\mathbf{R}\tau}\mathbf{M}\mathbf{q}_{\text{pss}}=\mathbf{P_D \Pi}.
\end{equation}
Using Eqs.~\ref{eq:stationary_dist_full_app} and \ref{stationary_longtime}, we get
\begin{equation}
\label{eq_qpss}
\begin{split}
     \mathbf{q}_{\text{pss}}^{\infty} &= \mathcal{N} 
 \bigg[\frac{2\kappa_L}{r} \ \frac{(\kappa_L + \kappa_R)}{r} \
 \frac{2\kappa_R}{r}  \  \frac{2}{r^2}  \
 \frac{2}{r^2}  \ \frac{2}{r} \ \frac{2}{r}  \ 2  \ 2 \bigg]^T,\\
 \mathcal{N}&=\frac{r^2}{ 4(1+r+r^2)+3r(\kappa_L + \kappa_R)},
 \end{split}
\end{equation}
where $\kappa_L=e^{-\beta  \left(\mu _L- \epsilon_0\right)}$, $\kappa_R=e^{-\beta  \left(\mu_R-\epsilon _0\right)}$, and $r=e^{-\beta \epsilon}$ with $\epsilon=(\epsilon_u-\epsilon_0)=(\epsilon_0-\epsilon_l)$.\par 

In the $\tau \to \infty$ limit, the circulation ($\Phi_{\infty}$) can be calculated using Eq.~\ref{circulation}.
The probabilities $p_{BE1}(0)$ and $p_{BE1}(\infty)$ are given by the $BE1$ elements of $\mathbf{M}\mathbf{q}_{\text{pss}}^{\infty}$ and $\mathbf{\Pi}$, respectively. Using Eqs.~\ref{work_def}, \ref{circulation}, \ref{stationary_state_full} and \ref{eq_qpss} we get

\begin{equation}
\label{work_long_time}
    W_{\infty}=\frac{\mathcal{N}(\mu_R-\mu_L)}{r}\left[\left(\frac{1+\delta}{2}\right)\kappa_R- \left(\frac{1-\delta}{2}\right) \kappa_L\right]
\end{equation}
Using Eqs.~\ref{stationary_state_full}, \ref{final_bit_dist} and \ref{delta_out}, we can describe the distribution of the outgoing bits as $p'_{0,1} =(1\pm\delta')/2$, where
\begin{equation}
\label{final_bit_dist_longtime}
    \delta' =\frac{r(\kappa_L - \kappa_R)}{4(1+r+r^2)+3r(\kappa_L + \kappa_R)},
\end{equation}
which can be used to calculate the entropy of the outgoing bits as $S'=-\sum_{i=0,1}p'_i\ln{p'_i} \in [0,\ln 2]$.

\subsubsection{Operational mode phase diagram}
\begin{figure*}
    \centering
    \includegraphics[scale=0.4]{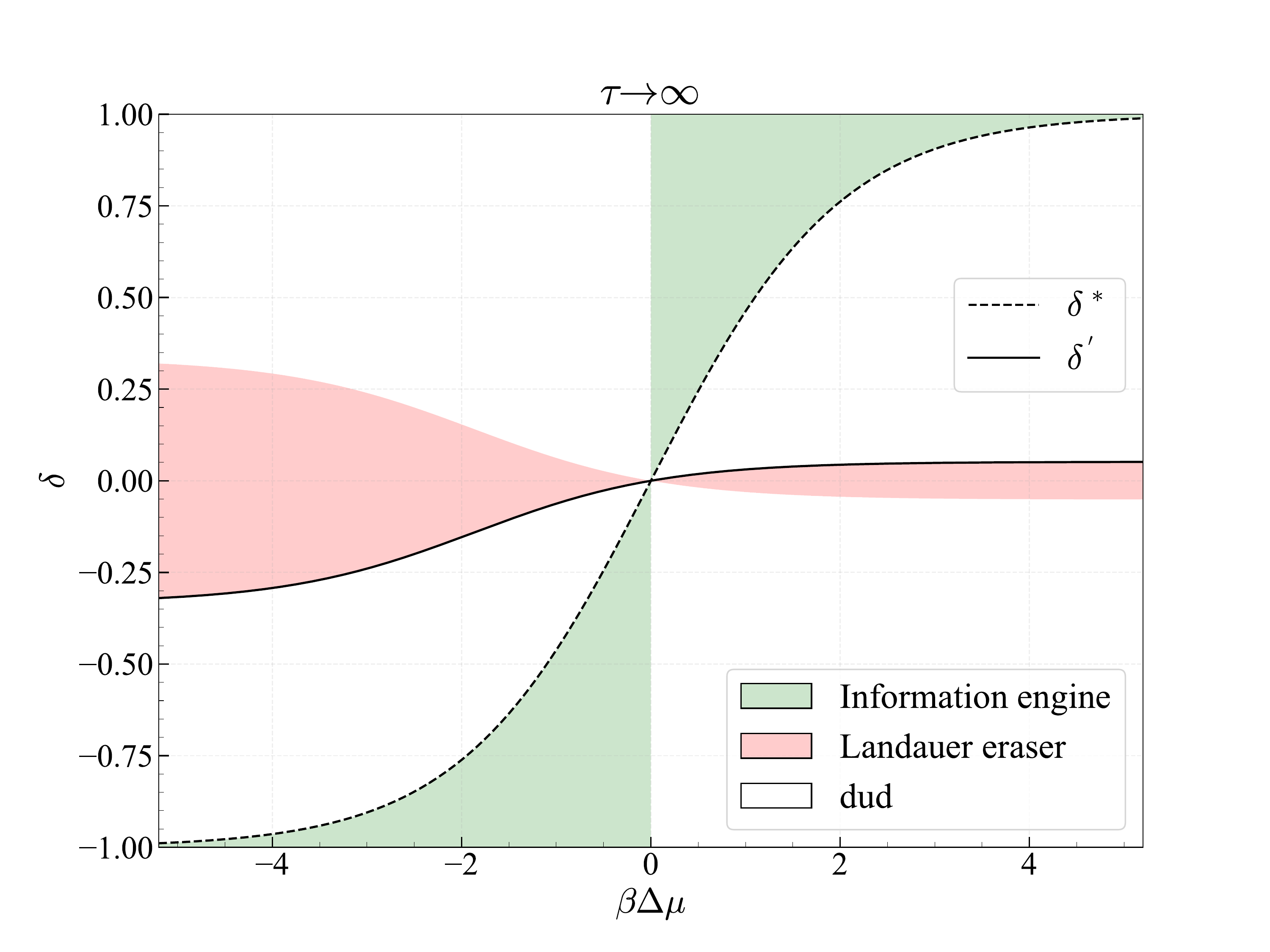}
    \caption{Analytically obtained phase diagram when $\tau \to \infty$. In the green region the system operates as an information engine ($\Delta S >0, W>0$) and in the red region it acts as a Landauer eraser. The critical parameter values $\delta^*$ and $\delta'$ are shown as function of $\beta \Delta \mu$ with $\Delta \mu = \mu_R - \mu_L$ with $\mu_L=0$. We have taken $\epsilon_0=0$, $\beta =1$ and $r=e^{-1}$ here.}
    \label{analytical_phase_diagram}
\end{figure*}

In the limit $\tau \to \infty$, bits in the outgoing bit-stream are uncorrelated and thus Eqs.~\ref{IPSL} and \ref{second_law} are equivalent, and both the final distribution $\delta'$ and the entropy of the outgoing bit become independent of $\delta$ (see Eq.~\ref{final_bit_dist_longtime}). The entropy change $\Delta S_\infty\equiv\lim_{\tau \to \infty} (S' -S)$ is a symmetric concave upwards function of $\delta$ with a negative value at its minimum ($\min_\delta \{\Delta S_{\infty}\}<0$) at $\delta=0$ when $\mu_L \ne \mu_R$.
Thus, in the region with $\abs{\delta}<\abs{\delta'}$ (shaded red in Fig.~\ref{fig:delS_and_W_compare_long_time_a}), we have $\Delta S_\infty<0$ and $ W_\infty<0$ (using Eq.~\ref{second_law}). By Eq.~\ref{final_bit_dist_longtime}, we see that when
\begin{equation}
    \abs{\delta}<\abs{\delta'}=\abs{\frac{r(\kappa_L - \kappa_R)}{4(1+r+r^2)+3r(\kappa_L + \kappa_R)}},
\end{equation}
information is erased from the incoming memory-tape, and the system consumes work, i.e., it acts as a Landauer eraser. Therefore, for a given value of $\Delta\mu=\mu_R-\mu_L$, the Landauer eraser region in the operational mode phase diagram is bounded by $\pm\delta'$, as indicated by the red regions in Fig.~\ref{analytical_phase_diagram}\par 

By Eq.~\ref{second_law}, $W_\infty>0$ implies $\Delta S_\infty>0$. Let $\delta^*$ denote the value of $\delta$ at which $\Phi_\infty=W_\infty/\Delta\mu$ changes its sign, for fixed $\mu_R$ and $\mu_L$. Using Eq.~\ref{work_long_time} we obtain
\begin{equation}
    \delta^*=\frac{\kappa_L-\kappa_R}{\kappa_L+\kappa_R}.
\end{equation}
Thus, $W_\infty>0$ when $\delta > \delta^*$ and $\mu_R>\mu_L$, or when $\delta<\delta^*$ and $\mu_R<\mu_L$. In these regions of parameter space, shown in green in Fig.~\ref{analytical_phase_diagram}, the system produces work at the cost of writing information to the memory-tape and the DQD acts as an information engine. \par 

In the regions of parameter space where $\Delta S_\infty > 0 >W_\infty $, information is written to the memory-tape and the system consumes work hence the model is a \textit{dud} \cite{MJ_model}.

\subsection{Semi-analytical results for finite $\tau$}
\label{Sec:phase_diagram}
For finite interaction time $\tau$, we can numerically diagonalize the transition rate matrix as $\mathbf{R}=\mathbf{UD_{R}V}$, where $\mathbf{D_{R}}$ is diagonal and $\mathbf{U}\mathbf{V}=\mathbf{V}\mathbf{U}=I$.
We then have
\begin{equation}
    \mathbf{T}= \mathbf{P_D }\mathbf{U}e^{\mathbf{D_R}\tau}\mathbf{V} \mathbf{M},
\end{equation}
and the evaluation of $e^{\mathbf{D_R}\tau}$ is straightforward.
Once $\mathbf{T}$ is obtained in this manner, the periodic steady state $\mathbf{q_{pss}}$ is calculated using Eq.~\ref{steady_state_eqn}, and thermodynamic quantities are determined as described in Sec.~\ref{Sec:Thermo}.

\begin{figure*}[t!]
   \begin{subfigure}[Phase Diagram ($\tau=0.2$)]{
    \includegraphics[scale=0.25]{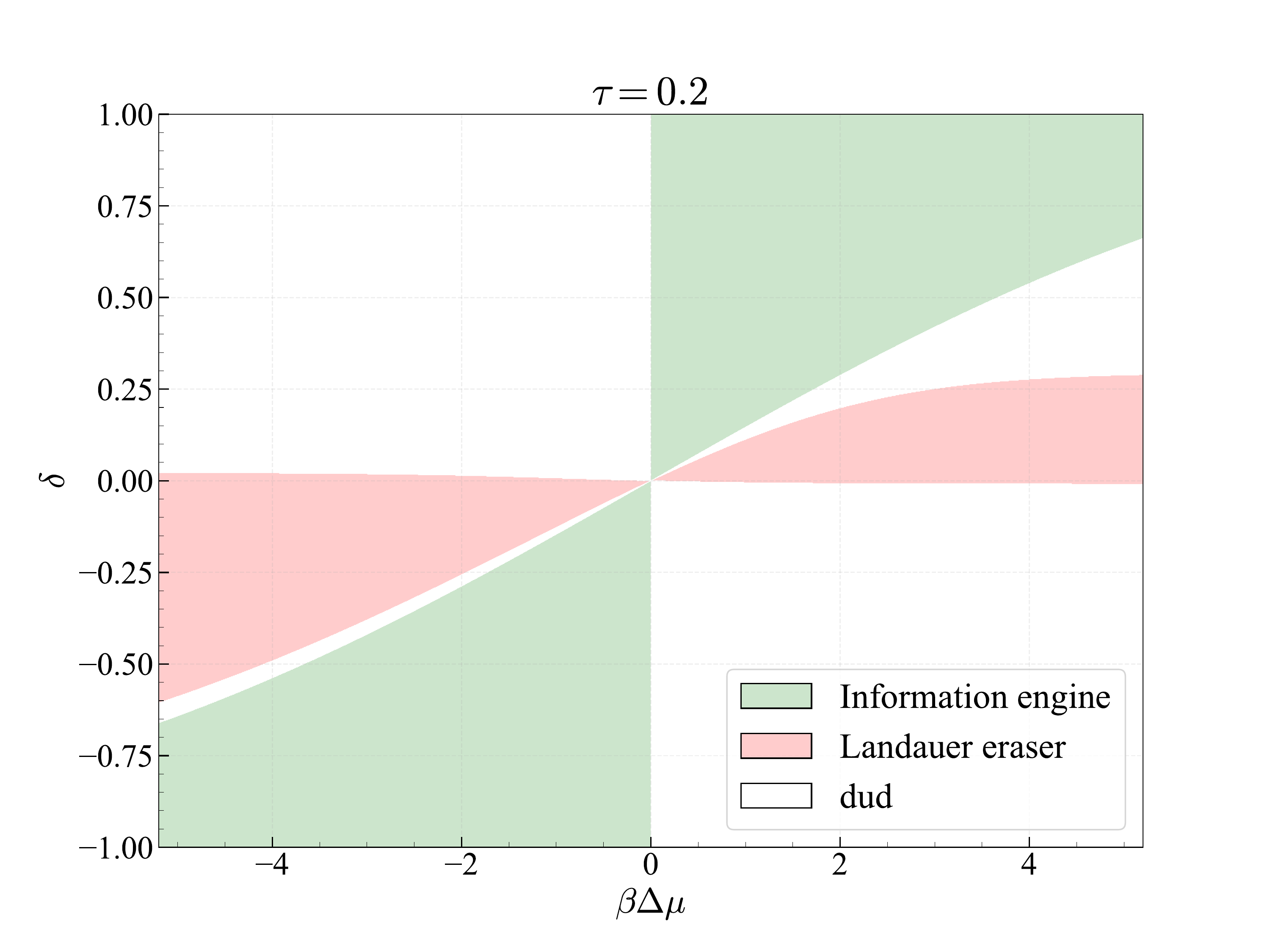}}
   \end{subfigure}
   \begin{subfigure}[Phase Diagram ($\tau=2.0$)]{
    \includegraphics[scale=0.25]{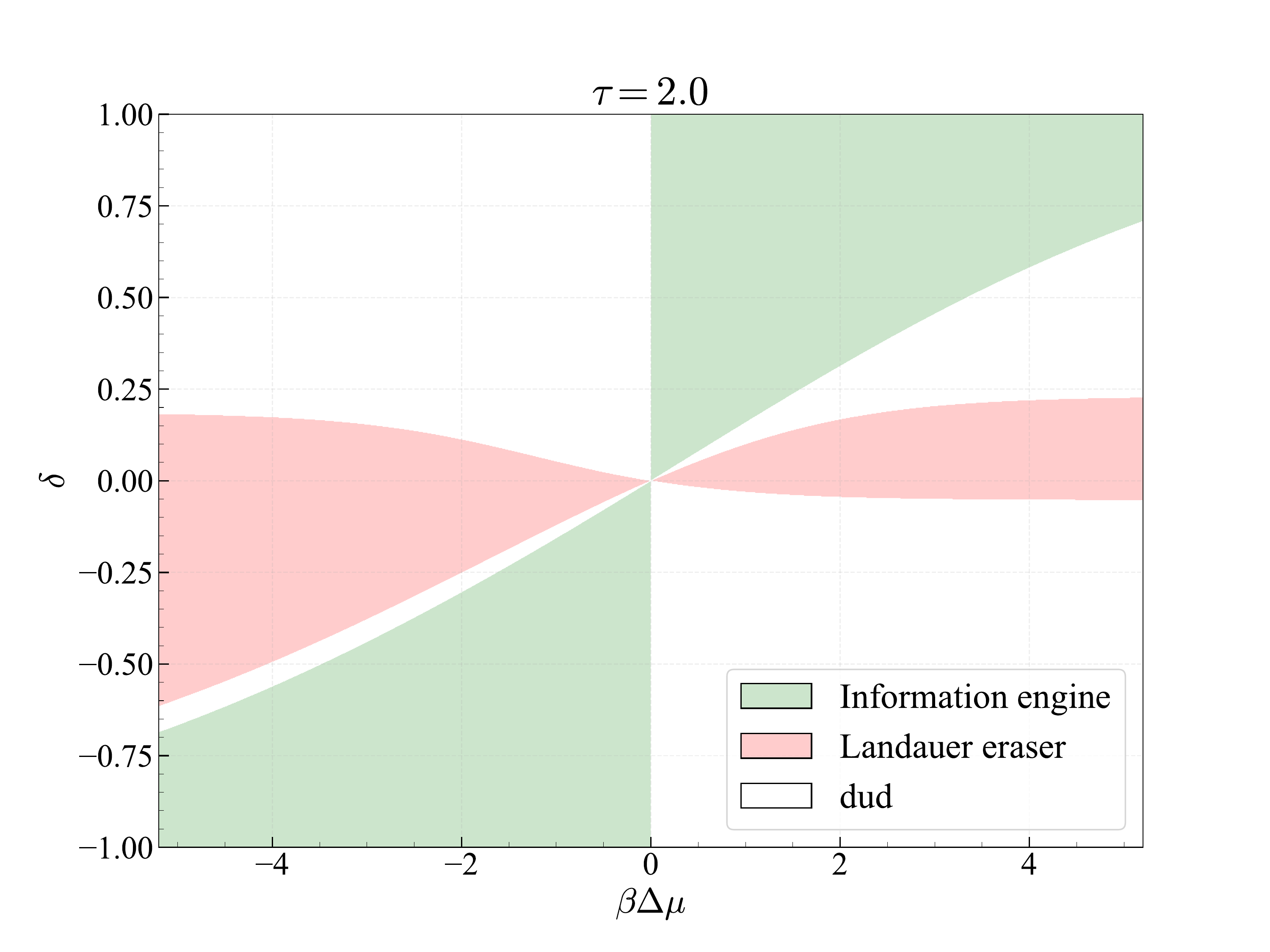}}
   \end{subfigure}
   \begin{subfigure}[Phase Diagram ($\tau=20.0$)]{
    \includegraphics[scale=0.25]{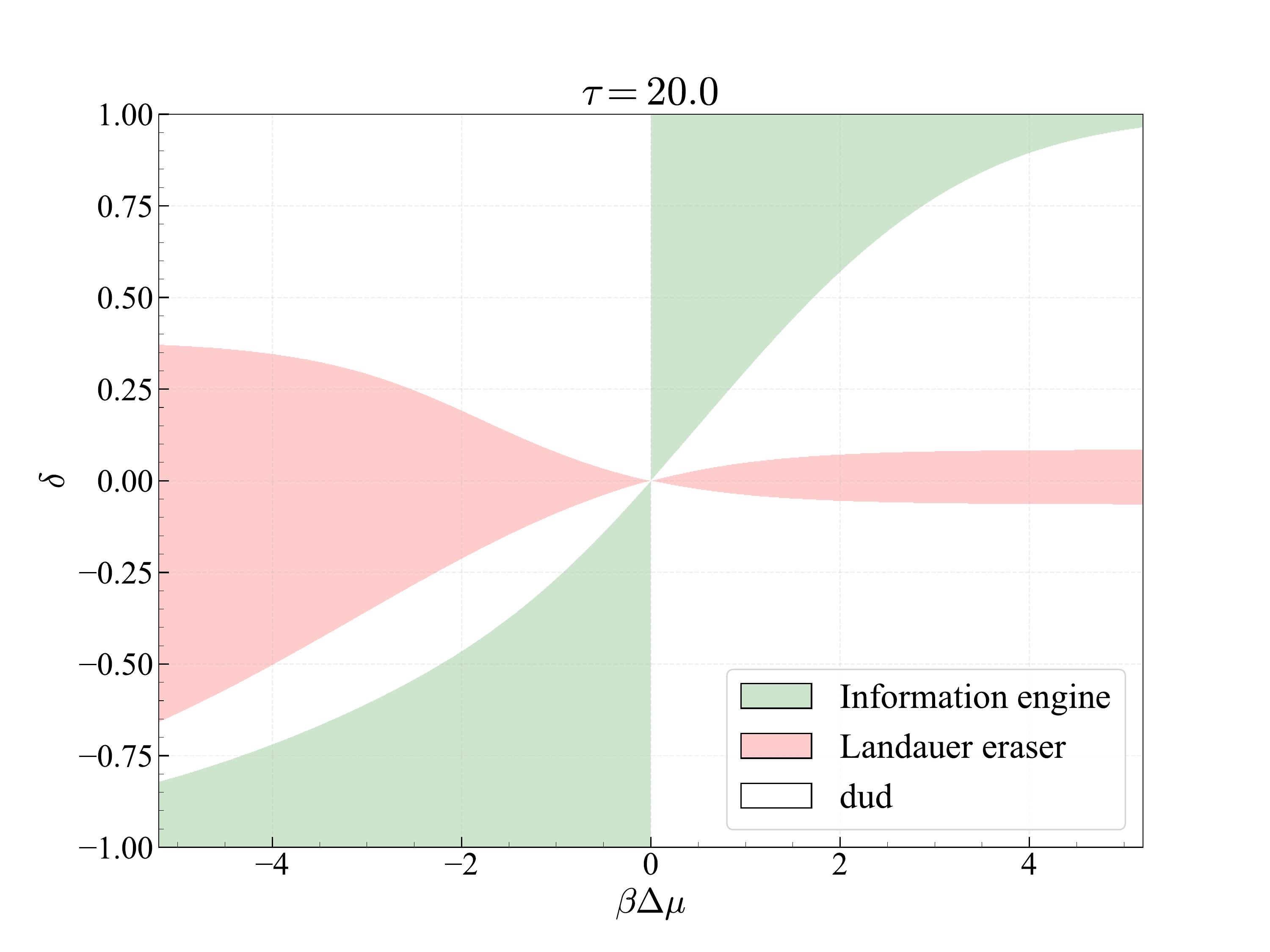}}
   \end{subfigure}
   \begin{subfigure}[Phase Diagram ($\tau=200.0$)]{
    \includegraphics[scale=0.25]{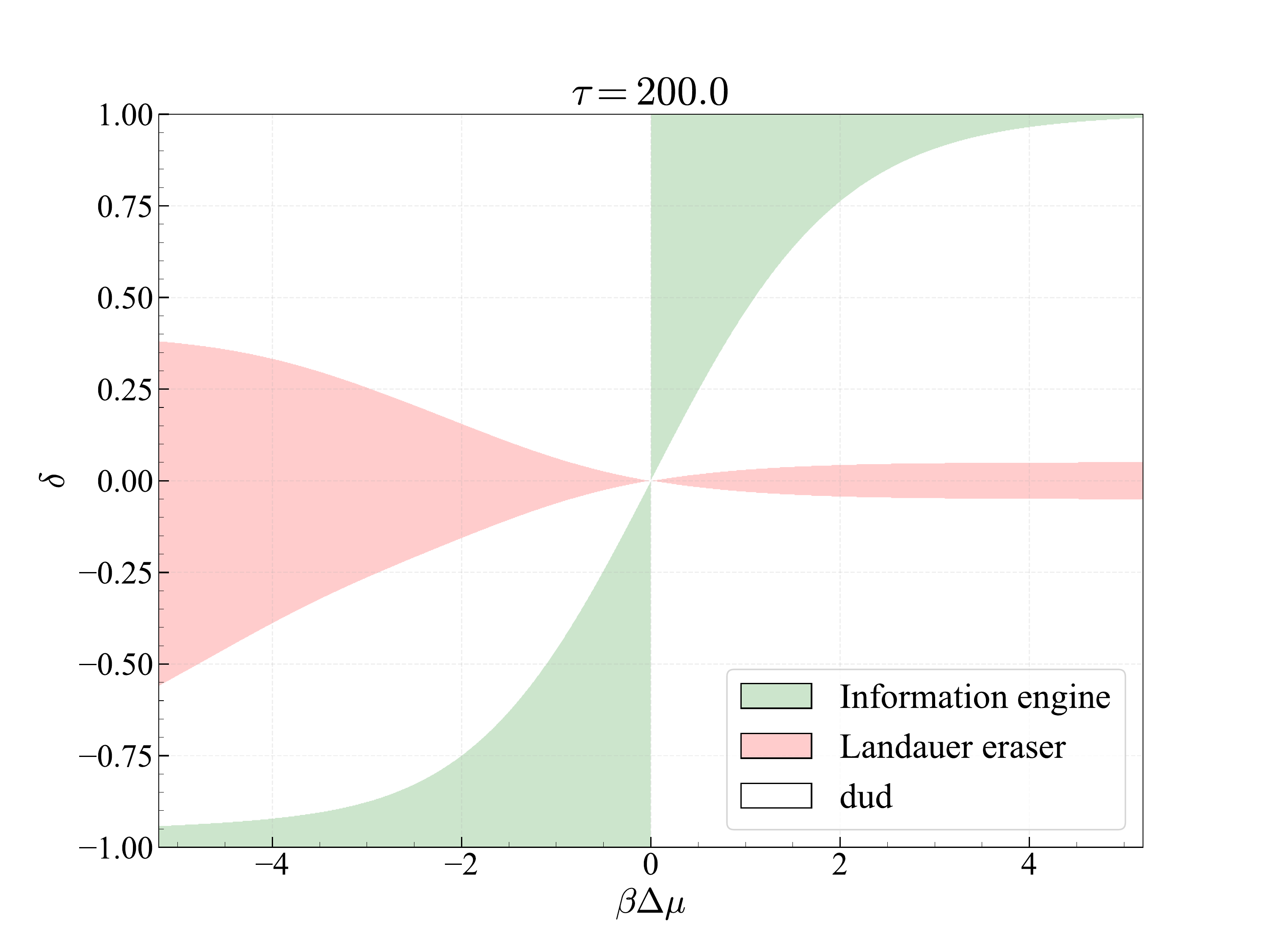}}
   \end{subfigure}
    \caption{Numerically obtained phase diagrams for different values of the interaction time, $\tau=0.2$, $2$, $20$ and $200$. For finite $\tau$, the final distribution of the memory-tape ($\delta'$) depends on the initial distribution ($\delta$), but this dependence vanishes in the limit $\tau \to \infty$. With increasing $\tau$, the phase diagram approaches the one shown in Fig.~\ref{analytical_phase_diagram}.
    We have fixed $r=e^{-1},\beta =1,\epsilon_0=0$.}
    \label{fig:finite_time_phase_diagram}
\end{figure*}
Following this semi-analytical approach, we have obtained phase diagrams for different values of $\tau$, using the second law inequality Eq.~\ref{second_law}, which is now the single symbol approximation to the IPSL in Eq.~\ref{IPSL}. 
Fig.~\ref{fig:finite_time_phase_diagram} shows these phase diagrams.
The competition between the effects of bit-coupling ($\delta$) and the thermodynamic bias ($\Delta \mu$) determines the direction of probability current, i.e, the sign of $\Phi$, in the network. 
With increasing values of $\tau$, the system has more time to relax to the equilibrium state $\mathbf{\Pi}$ before a new bit arrives, and the phase diagram approaches the one shown in Fig.~\ref{analytical_phase_diagram}.

In our model, the information engine region ($W >0$) appears only in the first and third quadrants of the phase diagrams. In these regions an increase in $\abs{\Delta \mu}$ increases the effective thermodynamic forces and suppresses the information engine region for a fixed value of $\delta$, as seen in Figs.~ \ref{analytical_phase_diagram} and \ref{fig:finite_time_phase_diagram}.\par 

For small values of $\tau$, the frequency of the virtual jumps in $\mathcal{G}_f$ (see Sec.~\ref{Sec:Energetics}) increases, as bits get replaced more frequently. These virtual jumps drive the probability current against the thermodynamic force in $\mathcal{G}_r$. Hence when $\tau$ is increased the information engine region decreases (see Fig.~\ref{fig:finite_time_phase_diagram}). \par

The entropy $S(\delta)=-\sum_{i}p_i\ln{p_i}$, with $p_{0,1}(\delta)=(1\pm\delta)/2$, is a concave downward function with a maximum at $\delta=0$.  As a result, when $\delta=0$ and $\delta'\ne 0$ we have $\Delta S =S(\delta')-S(\delta) <0$. This explains why the Landauer eraser region ($\Delta S<0$) contains the entire $\delta=0$ axis in the phase diagram (except for the origin $\delta=\beta\Delta\mu=0$, where $\Delta S=0$).

\subsection{Stochastic simulation}
\label{Sec:Stochastic_Sim}
\begin{figure*}[t!]
   \begin{subfigure}[Landauer eraser]{
      \label{fig:StochSimResults_a}
    \includegraphics[scale=0.25]{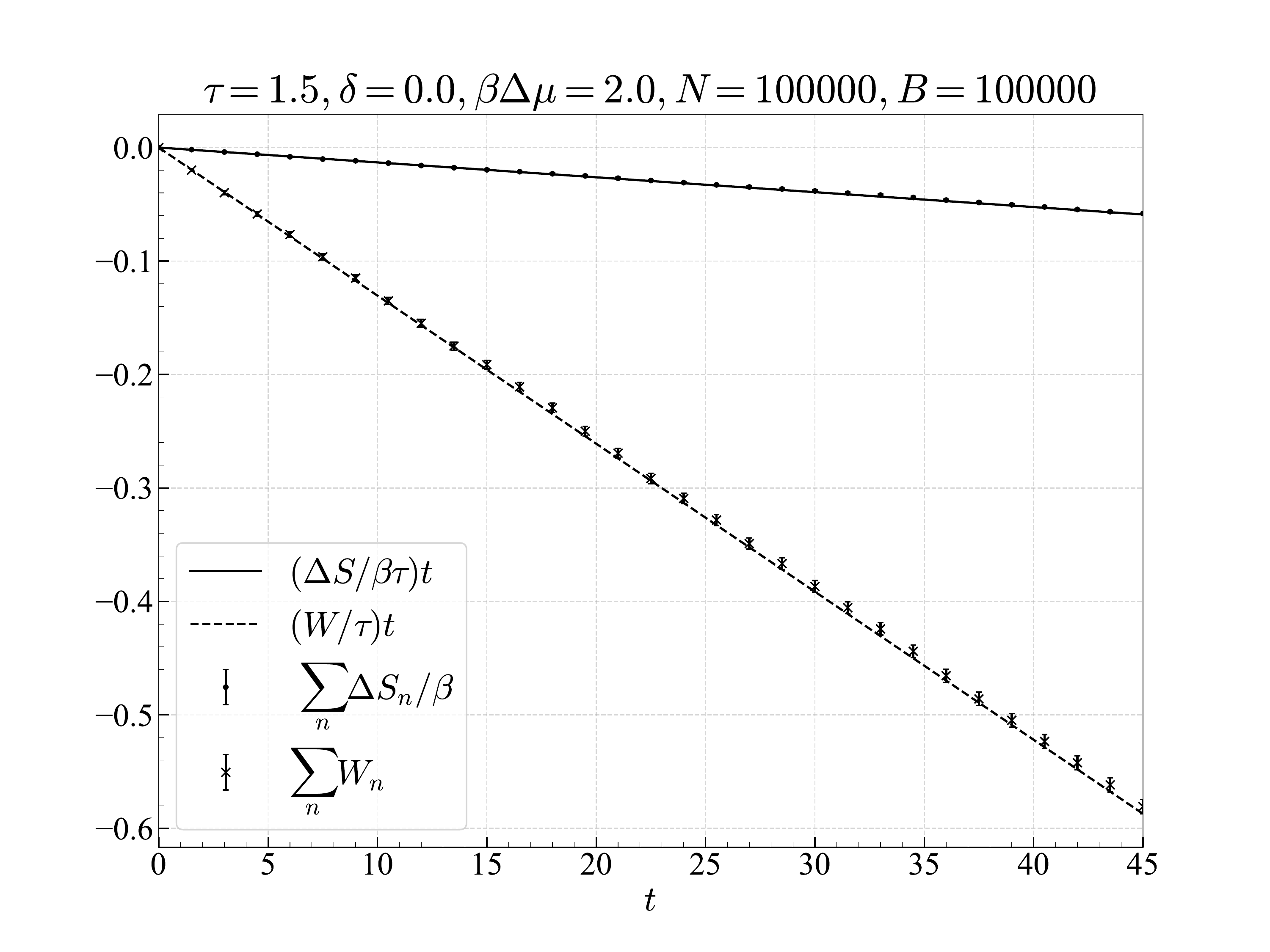}}
   \end{subfigure}
   \begin{subfigure}[Information engine]{
      \label{fig:StochSimResults_b}
    \includegraphics[scale=0.25]{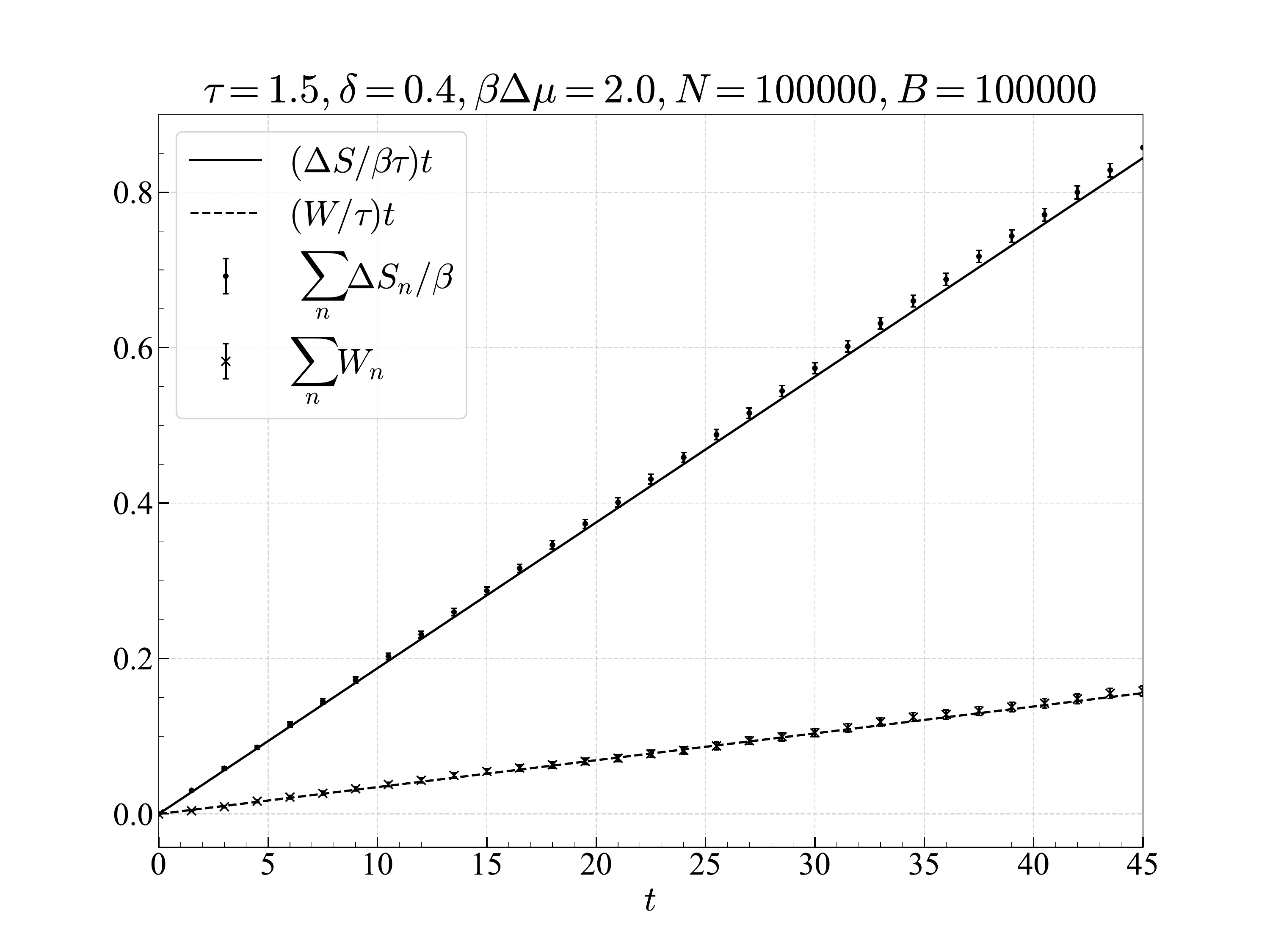}}
   \end{subfigure}
   \caption{Work and entropy production in (a) the Landauer eraser mode and (b) the information engine mode. Semi-analytical results and stochastic simulation results are compared. $\Delta S_n$ and $W_n$ represent the  change in single symbol entropy of the $n$th bit and average extracted work in the $n$th interval. For all the simulations, we have taken $\beta=1$, $r=e^{-1}$ and $\epsilon_0 = 0$. Errors are calculated with the bootstrap method. }
\end{figure*}
We have also performed stochastic simulations of the system. The variable $y=(\lambda,\sigma,b)$ was initialized by sampling $x=(\lambda,\sigma)$ from the distribution $\mathbf{q_{pss}}$, and $b$ from the distribution $\delta$.
During each bit interaction interval, $y$ evolves under a Poisson jump process, with the rates listed in Table~\ref{Ratematrixtable}.
At the end of each interval, the value of $b$ is replaced by the (randomly sampled) state of the incoming bit.
See Appendix \ref{appendix_stoch_sim} for further simulation details.

Figs.~\ref{fig:StochSimResults_a} and \ref{fig:StochSimResults_b} show work and entropy production when the system acts as a Landauer eraser and as an information engine, respectively. The total change in entropy $(\sum_n \Delta S_n)$ of the memory-tape was calculated by summing the change in single symbol entropy over each bit ($\Delta S_n$) in the memory-tape. Similarly, total work $(\sum_n W_n)$ was obtained by summing over the work done over each interval ($W_n$). In these figures, the semi-analytical results obtained by the approach described in Sec.~\ref{Sec:phase_diagram} are represented by straight lines with slopes $\Delta S/\tau$ and $W/\tau$. $N=10^5$ trajectories were generated, and statistical errors in $\Delta S_n$ and $W_n$ were calculated using the bootstrap method~\cite{efron1994introduction}, by resampling $B=10^5$ times with replacements~\cite{efron1994introduction}. The increasing errors in $\sum_n \Delta S_n $ and $\sum_n W_n$ reflect the accumulation of statistical errors with each additional interaction interval.

\section{Conclusion}
\label{Sec: Conclusion}
We have presented a strategy for constructing a memory-tape model of Maxwell's demon, from a feedback-controlled model.
We have illustrated this strategy using the Annby-Andersson model \cite{AA_model}, a feedback-controlled Maxwell's demon in a double quantum dot (DQD). In our approach, we replace the feedback controller with a stochastic variable evolving under the same thermal environment as the DQD. We then couple our system to an information reservoir and design suitable bit interaction rules to mimic the effects of the feedback controller. In analyzing our model, we obtained an exact solution in the limit of infinitely long interaction time $\tau$, and used a semi-analytical approach involving numerical matrix diagonalization for finite $\tau$. As illustrated by these results as well as stochastic simulations, our model can act both as an information engine and as a Landauer eraser, for suitable parameter values.

Our research strengthens the connection between two paradigms of information thermodynamics: 
Maxwell's original, non-autonomous paradigm of a ``nimble-fingered'' demon performing feedback control at the level of thermal fluctuations; and the autonomous paradigm, due to Bennett~\cite{Bennet1982}, in which the demon is replaced by a physical gadget, thermodynamically driven by the continual randomization of a stream of bits (the memory-tape).
In effect, given a demon, we show how to design a gadget that mimics it.

Our approach makes use of the underlying network structure of a feedback-controlled system, and it relates to recent stochastic thermodynamic studies of bipartite systems \cite{Bipartite_2014,Hartich_Barato_Seifert_Bipartite_2014,Diana_Esposito_Bipartite_2014}. Specifically, the dynamics of $y=(\lambda,\sigma, b)$ in $\mathcal{G}_f$ can be described as bipartite system dynamics by splitting $y$ in two random variables: $\sigma$ and
$\Bar{x}\equiv(\lambda,b)$ that do not change simultaneously. This is in contrast with the original MJ model \cite{MJ_model} which lacks the bipartite structure (see \cite{Bipartite_2014}). 

Double quantum dot systems are promising candidates for the experimental implementation of information engines \cite{Barker2019}. We note that while there have been a number of realizations of feedback-controlled demons \cite{Rex_review, Parrando_review}, experimental realizations of memory-tape models are yet to be explored.
By showing how to design a memory-tape model that mimics a feedback controlled system, our approach may be useful in the design of physical implementations of autonomous information engines.

Although our analysis has been entirely at the level of classical stochastic dynamics, it would be worth studying analogous quantum models (see e.g.~\cite{Deffner_quantum}).  A future research direction might explore design principles for quantum analogs of the memory-tape model.
Lastly, we limited our discussion of the information-theoretic aspects of this model to the single symbol entropy. The study of the effects of correlations among the bits offers another avenue for future research.
\section*{Acknowledgements}
This research was supported by
grant number FQXi Grant Number: FQXi-IAF19-07
from the Foundational Questions Institute Fund, a donor
advised fund of Silicon Valley Community Foundation. We thank Guilherme
De Sousa, Bj\"{o}rn Annby-Andersson, Faraj Bakhshinezhad, Peter Samuelsson and Patrick P. Potts for fruitful discussions. 
\appendix
\section{Rate matrix and unique stationary state for $\mathcal{G}_f$}
\label{app:RateMatrix}
The rate matrix $\mathbf{R}$ and stationary distribution $\mathbf{\Pi}$ are
\begin{widetext}
\begin{equation}
\label{eq:rate_matrix_full}
\mathbf{R}=
\left(
\begin{array}{cccccccccccccccccc}
 \bm{-1} & 0 & 0 & 0 & 0 & \bm{\kappa_L} & 0 & 0 & 0 & 0 & 0 & 0 & 0 & 0 & 0 & 0 & 0 & 0 \\
 0 & \bm{-1} & 0 & 0 & 0 & 0 & 0 & 0 & 0 & \bm{1} & 0 & 0 & 0 & 0 & 0 & 0 & 0 & 0 \\
 0 & 0 & \bm{-2} & 0 & 0 & 0 & \bm{\kappa_R} & 0 & 0 & 0 & \bm{1} & 0 & 0 & 0 & 0 & 0 & 0 & 0 \\
 0 & 0 & 0 & \bm{K_1} & \bm{1} & 0 & 0 & 0 & \bm{1} & 0 & 0 & 0 & 0 & 0 & 0 & 0 & 0 & 0 \\
 0 & 0 & 0 & \bm{1} & \bm{K_2} & 0 & 0 & \bm{1} & 0 & 0 & 0 & 0 & 0 & 0 & 0 & \bm{1} & 0 & 0 \\
 \bm{1} & 0 & 0 & 0 & 0 & \bm{K_3} & 0 & \bm{1} & 0 & 0 & 0 & 0 & \bm{r} & 0 & 0 & 0 & 0 & 0 \\
 0 & 0 & \bm{1} & 0 & 0 & 0 & \bm{K_4} & 0 & \bm{1} & 0 & 0 & 0 & 0 & 0 & 0 & 0 & 0 & 0 \\
 0 & 0 & 0 & 0 & \bm{r^2} & \bm{r} & 0 & \bm{-2} & 0 & 0 & 0 & 0 & 0 & 0 & 0 & 0 & 0 & 0 \\
 0 & 0 & 0 & \bm{r^2} & 0 & 0 & \bm{r} & 0 & \bm{-2} & 0 & 0 & 0 & 0 & 0 & 0 & 0 & 0 & 0 \\
 0 & \bm{1} & 0 & 0 & 0 & 0 & 0 & 0 & 0 & \bm{-2} & 0 & 0 & 0 & 0 & \bm{\kappa_L} & 0 & 0 & 0 \\
 0 & 0 & \bm{1} & 0 & 0 & 0 & 0 & 0 & 0 & 0 & \bm{-1} & 0 & 0 & 0 & 0 & 0 & 0 & 0 \\
 0 & 0 & 0 & 0 & 0 & 0 & 0 & 0 & 0 & 0 & 0 & \bm{-1} & 0 & 0 & 0 & \bm{\kappa_R} & 0 & 0 \\
 0 & 0 & 0 & 0 & 0 & \bm{1} & 0 & 0 & 0 & 0 & 0 & 0 & \bm{K_2} & \bm{1} & 0 & 0 & 0 & \bm{1} \\
 0 & 0 & 0 & 0 & 0 & 0 & 0 & 0 & 0 & 0 & 0 & 0 & \bm{1} & \bm{K_1} & 0 & 0 & \bm{1} & 0 \\
 0 & 0 & 0 & 0 & 0 & 0 & 0 & 0 & 0 & \bm{1} & 0 & 0 & 0 & 0 & \bm{K_5} & 0 & \bm{1} & 0 \\
 0 & 0 & 0 & 0 & \bm{r} & 0 & 0 & 0 & 0 & 0 & 0 & \bm{1} & 0 & 0 & 0 & \bm{K_6} & 0 & \bm{1} \\
 0 & 0 & 0 & 0 & 0 & 0 & 0 & 0 & 0 & 0 & 0 & 0 & 0 & \bm{r^2} & \bm{r} & 0 & \bm{-2} & 0 \\
 0 & 0 & 0 & 0 & 0 & 0 & 0 & 0 & 0 & 0 & 0 & 0 & \bm{r^2} & 0 & 0 & \bm{r} & 0 & \bm{-2} \\
\end{array}
\right),
\end{equation}
\begin{equation}
    \label{eq:stationary_dist_full_app}
    \begin{split}
    \mathbf{\Pi}&= \frac{r^2}{ 4(1+r+r^2)+3r(\kappa_L + \kappa_R)}
    \left[
    \begin{array}{c}
        \kappa_L r^{-1} \\ \kappa_L r^{-1} \\ \kappa_R r^{-1} \\ r^{-2}  \\ r^{-2} \\ r^{-1} \\ r^{-1} \\ 1 \\ 1 \\
        \kappa_L r^{-1} \\ \kappa_R r^{-1} \\ \kappa_R r^{-1} \\ r^{-2} \\ r^{-2} \\ r^{-1} \\ r^{-1} \\ 1 \\ 1
        \end{array}
        \right]
    \end{split}
\end{equation},
\end{widetext}
where $\kappa_L=e^{-\beta  \left(\mu _L- \epsilon_0\right)}$, $\kappa_R=e^{-\beta  \left(\mu_R-\epsilon_0\right)}$, $K_1=-r^2 -1$, $K_2=-r^2-r-1$, $K_3 = -\kappa_L-r-1$, $K_4=-\kappa_R-r$, $K_5=-\kappa_L-r$ and $K_6=-\kappa_R-r-1$. Here the states in $V(\mathcal{G}_f)$ are ordered as follows: ($AE0$, $BE0$, $CE0$, $BL0$, $BR0$, $AL0$, $CR0$, $AR0$, $CL0$, $AE1$, $BE1$, $CE1$, $BL1$, $BR1$, $AL1$, $CR1$, $AR1$, $CL1$).

\section{Details of stochastic simulation scheme}
\label{appendix_stoch_sim}
\subsubsection{Poisson jumps } We implement the Gillespie Algorithm \cite{Gillespie_review,GILLESPIE_1976_JCompPhys,Gillespie_JPC_1977} to simulate
the Poisson jump process for $y$ in $\mathcal{G}_f$, when the DQD system is interacting with a bit. If a system is in state $y_j$ at time $t$, then the time interval for the next jump event is generated from the Poisson distribution as follows:
\begin{equation}\label{MC-step_time_interval}
    \Delta t= \frac{1}{\sum_{y\ne y_j}R_{y y_j}}\ln{\frac{1}{\xi_1}}
\end{equation}
where $\xi_1$ is sampled uniformly in the interval $(0,1]$. After remaining in the state $y_j$ over the time interval $[t,t+\Delta t)$, the system jumps to a new state (say $y_{j'}$). To find $y_{j'}$, all states in $V(\mathcal{G}_f)$ are arranged in order (say, $(0,1,2,..,16,17$)) then $j'$ is chosen as the smallest integer label of the ordered states that satisfies:
\begin{equation}\label{MC-step-Gillespie_update}
    \frac{\sum_{i=0,y_i \ne y_j}^{j'}R_{y_iy_j}}{\sum_{y\ne y_j}R_{y y_j}} > \xi_2
\end{equation}
where $\xi_2$ is sampled uniformly in the interval $(0,1]$.

\subsubsection{Virtual jumps}
\label{Sec: Stoch_Sim_virtual_jumps}
Virtual jumps occur when a new bit arrives. 
Specifically, if $y=(x_j,b_n)$ at time $t \in (n\tau,(n+1)\tau)$, and if $t+\Delta t > (n+1)\tau$, then instead of generating a jump using Eq.~\ref{MC-step-Gillespie_update}, a new bit state is generated at time $(n+1)\tau$.

The new incoming bit is sampled with probability $p_0$ ($p_1$) to be in state $b_{n+1}=0$
($b_{n+1}=1$), the state $y$ is updated to $y_{j'}\equiv (x_j,b_{n+1})$, and the time is set to $t=(n+1)\tau$. We express this update rule as
\begin{eqnarray}
    &b&_{n+1} =
    \begin{cases}
    0, \, \text{with probability} \,\, p_0 \\ 
    1, \, \text{with probability} \,\, p_1
    \end{cases} \\
    &y&((n+1)\tau) = (x_j,b_{n+1})
\end{eqnarray}
when $t+\Delta t > (n+1) \tau$.
If $b_{n+1}\ne b_n$ then this update constitutes a virtual jump, otherwise the state of $y$ is unchanged.

\bibliography{main}

\end{document}